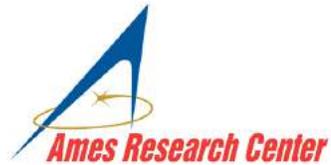 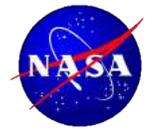

# TECHNOLOGY DEVELOPMENT FOR EXOPLANET MISSONS

Technology Milestone Final Report

*Enhanced direct imaging exoplanet detection with astrometry mass determination*

October 09, 2017

Eduardo Bendek, P.I.
Ruslan Belikov, Olivier Guyon,
Thomas Greene, Eugene Pluzhnik, Tom Milster,
Alexander Rodack, Emily Finan, and Justin Knight.

National Aeronautics and Space Administration
Ames Research Center
Moffett Field, California

# Signature page

**Approvals:**

Eduardo Bendek, Principal Investigator                                            Date
NASA Ames Research Center/BAERI

Nicholas Siegler, Program Chief Technologist                                      Date
Exoplanet Exploration Program, NASA / JPL - California Institute of Technology

Brendan Crill, Deputy Program Chief Technologist                                  Date
Exoplanet Exploration Program, NASA / JPL - California Institute of Technology

Douglas Hudgins, Program Scientist                                                Date
Exoplanet Exploration Program, Science Mission Directorate, NASA HQ

# 1  Executive summary

This Final Report (FR) presents the results of the *Enhanced direct imaging exoplanet detection with astrometry mass determination* project, which was executed in support of NASA's Exoplanet Exploration Program and the ROSES Technology Development for Exoplanet Missions (TDEM). This FR also provides specific information about milestone compliance, methodology for computing its metrics, and establishes the success criteria against which the milestone will be evaluated. The first milestone is concerned with a demonstration of medium fidelity astrometry accuracy and the second milestone demonstrates high-contrast imaging utilizing the same astrometry-capable optics.

The scientific importance of measuring planet masses increases as the exoplanet community focuses on exoplanet characterization necessary to answer the questions stated in the NASA's strategic documents such as the "Enduring Quests, Daring Visions" 30-year roadmap and the New Worlds New Horizons in Astronomy and Astrophysics (NWNH) Astro2010 decadal survey. In fact, about one-third of the 30-year roadmap considers the question, "Are We Alone?" In the Astro2010 decadal survey, one of the key questions posed is "Do habitable worlds exist around other stars?" More specifically, the NWNH also envisions the "*search for nearby, habitable, rocky or terrestrial planets with liquid water and oxygen…*" in the 2020 Vision chapter.

Also, the NWNH states the importance of studying nearby stars, "*Stars will then be targeted that are sufficiently close to Earth that the light of the companion planets can be separated from the glare of the parent star and studied*" (pg. 39 paragraph 1). This target distance regime is where combined direct imaging and astrometry is most efficient because the signal scales inversely proportional to the target distance.

More specifically, masses are particularly important in the study of earth-like planets around sun-like stars to be able to distinguish them from small Neptunes or water worlds. Masses are essential to resolve many potential ambiguities in interpretations of images or spectra of observed planets. This in turn informs the phase of the imaging observation, which informs the planet's albedo and its atmospheric composition. Masses also constrain the planet's surface gravity, essential for understanding the spectra of gaseous planet atmospheres. Future flagship missions with limited number of targets, such as HABEX and LUVOIR, could greatly benefit from having a precision astrometry capability.

The main factor that limits the accuracy of sparse-field astrometry, aside from photon noise, is dynamic distortion that arises from perturbations in the optical train. Even space optics suffer from dynamic distortions in the optical system at the sub-$\mu$as level. The work carried out during this technology development grant has allowed us to demonstrate and advance the Diffractive Pupil (DP) technology used to calibrate optical system distortions. As a result, we are paving the way to achieve stellar astrometry accuracy of sub-$\mu$as levels that would enable mass measurements of earth-like planets around nearby Sun-like stars. In addition, we have shown that the DP technology is technically fully compatible and scientifically synergistic with a coronagraph.

In the scope of this funded effort, we have designed, built, and tested an optical bench capable of performing simultaneous direct imaging and astrometry measurements to demonstrate the compatibility of both techniques. Since every optical surface in the system produces some contribution to the system distortion, we placed the DP on the first optical element to calibrate the entire optical system. As a result, the DP is placed on the secondary mirror of the "collimating telescope" which also serves as the stop and tip/tilt control for the system. The collimated output feeds a second telescope, which serves as a camera and creates an image of the sources on the astrometry camera. A small pick-off mirror is placed close to the focal plane to extract the central star light and send it to a coronagraph instrument. We use a PIAA coronagraph to apodize the beam and a Deformable Mirror to perform speckle nulling.

To run the system, we utilized heaters with closed-loop control to maintain the system temperature at 1°C above ambient. The system temperature remained stable within +/- 20mK during typical 48-hour data runs.

## 1.1 Major accomplishments and milestone compliance

The milestone #1 for this TDEM effort is about performing a broadband medium fidelity imaging astrometry, and it is defined as follows: *Demonstrate $2.4 \times 10^{-4}$ $\lambda/D$ astrometric accuracy per axis performing a null result test. The laboratory work will be carried out in broadband spectrum covering wavelengths from 450 to 650nm using an aperture pupil (D) equal or larger than 16mm.*

We have met milestone #1 with comfortable margin. The average accuracy obtained over the three data sets is $5.75 \times 10^{-5}$ $\lambda/D$, which is 4 times better than the milestone requirement, or equivalent to $2.5\mu as$ on 2.4m telescope, or $1.5\mu as$ for a 4m telescope, working in visible band. These results show the potential of this technique to enable detection and measure masses of earth-like planets around nearby stars, hence bringing a real benefit to the astronomy community.

The milestone #2 for this TDEM effort is about broadband medium fidelity simultaneous imaging astrometry and high-contrast imaging, and it is defined as follows: *Demonstration of milestone #1, and performing high-contrast imaging achieving $5 \times 10^{-7}$ raw contrast between 1.6 and $6\lambda/D$ by a single instrument, which shares the optical path, from the source to the coronagraphic and astrometry field of view (FoV) separation. The ability of achieving $5 \times 10^{-7}$ raw contrast will be considered as proof of no contamination of the IWA.*

We also met milestone #2 and demonstrated that it is possible to achieve high-contrast imaging utilizing a coronagraph fed by a telescope equipped with a DP, enabling dual use of the telescope. We performed three different high-contrast imaging runs and met the milestone #2 of $5 \times 10^{-7}$ raw contrast for all of them. On average, we obtained $3.33 \times 10^{-7}$ raw contrast considering all data sets. This result is 35% better than the milestone #2 requirement. We validated the stability of the high-contrast region by averaging frames and subtracting the average from single frames, which resulted in contrast improvement of approximately one order of magnitude, reaching $2.72 \times 10^{-8}$ contrast.

The main achievement of this work was the medium fidelity demonstration and feasibility validation of performing astrometry and direct imaging using the same instrument, significantly enhancing the expected scientific yield of dedicated exoplanet characterization missions.

## 1.2 Secondary accomplishments

During the development of this TDEM, we advanced several technologies including dot imprinting on curved substrates done together with University of Arizona, which enabled photolithography on a gimbal mount and precise pattern stitching.

Also, we tested the novel Miniaturized Deformable Mirror controllers, which we developed at Ames as part of a Center Innovation Fund grant. We used it on the device manufactured for this TDEM project, proving the reliability, stability and performance of this unique technology.

# 2 Table of Contents

# 3 TDEM description

## 3.1 TDEM Goals

This work aims to demonstrate the feasibility of combining direct imaging and stellar astrometry, [Guyon et al 2013a] validating the scientific and cost advantages of the approach by developing an integrated laboratory for testing and performance characterization. We aim to support NASA strategic plans by informing future mission concept design teams with the state-of-the-art of this technology. We expect to provide new knowledge to advance the following specific goals as a result of the milestone completion:

- Demonstrate in the laboratory that direct imaging and precision astrometry can be performed with the same telescope equipped with a DP [Guyon et al 2012a]. This will enable a single exoplanet flagship mission to perform the task that was originally expected to require two separate missions (high-precision astrometry and direct imaging), resulting in substantial advantages in cost and schedule.
- Understand the technical implementation challenges of combining coronagraphy direct imaging and DP astrometry in greater depth to generate a detailed error budget needed for a future high fidelity demonstration and mission concept planning.
- Explore the potential benefits of using the astrometry signal to accurately and independently measure spacecraft pointing, characterize flexing of the spacecraft bus and optical components by means of tomographic reconstruction, using the motion of the diffractive spikes.
- Advance mirror coating technology that allows applying high performance coatings with special shapes needed for astrometry and other advanced optical calibrations [Bendek el al. 2013c].

## 3.2 Laboratory overview

The combined astrometry and direct imaging laboratory developed for this TDEM, which will be referred as Astrometry Demonstration (AD) for the rest of the document, is located inside the Ames Coronagraph Experiment (ACE) Laboratory [Belikov et al 2010] at the NASA Ames Research Center (ARC). The AD was installed inside the ACE lab class ~100,000 clean room, next to the ACE lab optical bench. Both experiments benefit each other by sharing a variety of state-of-the-art instrumentation that includes two Kilo Deformable Mirrors, Phase Induced Amplitude Apodization (PIAA) [Guyon et al. 2003] lenses and mirrors, a Zygo interferometer, and a broadband supercontinuum laser. The AD optical system is built on top of a 3'x 6' Thorlabs optical breadboard, which is supported by four passive isolation optical table supports. Control electronics, data loggers, lab computer, laser sources and a cooling chiller are mounted away from the optical bench to avoid vibrations and thermal loads.

The optical bench consists of an imaging system from which the on-axis field is extracted to feed a high-contrast imaging instrument. The remaining off-axis-field is imaged by a wide-field instrument, which delivers the images for astrometry measurement. As a result, our experiment is an end-to-end test bed of a simultaneous direct imaging and astrometry mission.

## 3.3 System optical layout

The system starts with a ***light source***, which has an array of 21x21 sources, spaced every 1mm, and arranged on a squared grid, defining an object of 20x20 mm size. The central source is about four orders of magnitude brighter, and serves as the central star. A super continuum laser manufactured by NKT Photonics, model *SuperK EXTREME*, is used to feed white laser light to the source.

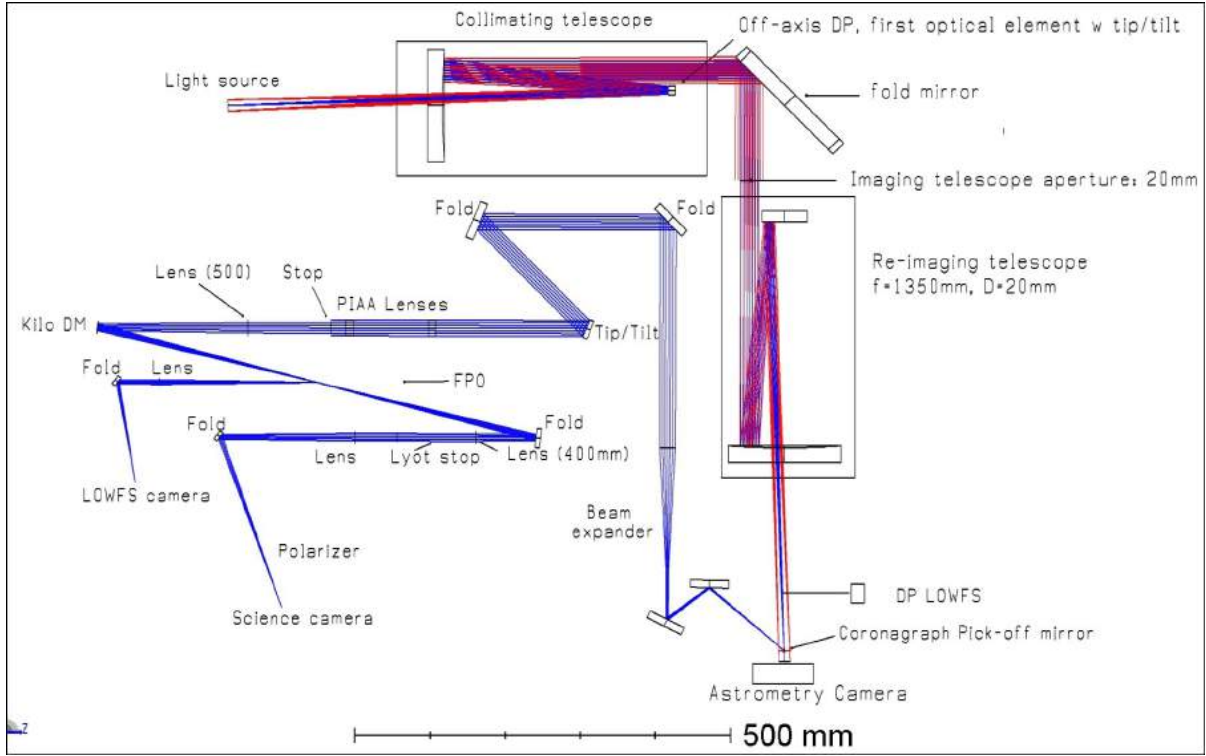

Figure 3.1. Astrometry Demonstration optical layout showing the light source, the collimating and imaging telescopes, and the coronagraph on the left side.

The sources are reimaged at infinity using a ***collimator telescope,*** which is an off-the-shelf 6" Ritchey-Chretien (RC) f/9 telescope with a focal length of 1370mm. The telescope is reverse-illuminated from its focal plane as shown in Fig 3.1.

The collimator telescope's secondary mirror, shown in Fig 3.2, is the first optical surface of the system; hence the DP is imprinted on its surface. We modified the telescope secondary to add tip/tilt focus actuation using hybrid lead-screw/piezo Thorlabs PE-4 actuators. These actuators have a stroke range of 15μm when 150V are applied, delivering an average 0.1μm per volt. The actuators are 30mm apart from each other resulting on 0.69"/V tilt in the secondary. This tilt rate translates into 4.56μm/V motion on the astrometry focal plane.

To avoid diffraction from the telescope spiders, we use an unobstructed subaperture stop with a diameter of 20mm and an off-axis distance of 20mm, as shown in Fig. 3.1. The red rays represent the 18.6' HFoV and the blue is the on-axis beam. The collimated off-axis beam that exits the telescope is folded and directed to the ***imaging telescope***, which is another 6" f/9 RC telescope. Before entering its

Table 3.1: Astrometry experiment design parameters

| Design parameters | Angular | Focal plane |
|---|---|---|
| Wavelength range | 400 – 690nm | N/A |
| Focal length | 1.37m | N/A |
| Aperture D | 0.020m | N/A |
| Sampling factor | 3 | N/A |
| Pixels/(L/D) | 6.1px | 45.1μm |
| Pixel size | 0.165λ/D | 7.4μm |
| Detector size | 2000 px | 14.8mm |
| HFoV | 18.6' | 7.4mm |
| HFoV | 165 λ/D | 7.4mm |
| **Diffractive Pupil** | **On mirror** | **Pupil scaled** |
| DP spacing *a* | 120μm | 240μm |
| Dot diameter | 12μm | 24μm |
| DP diameter | 10mm | 20mm |
| 1st Order location | 55 λ/D | Same |
| 2nd Order location | 95 λ/D | Same |

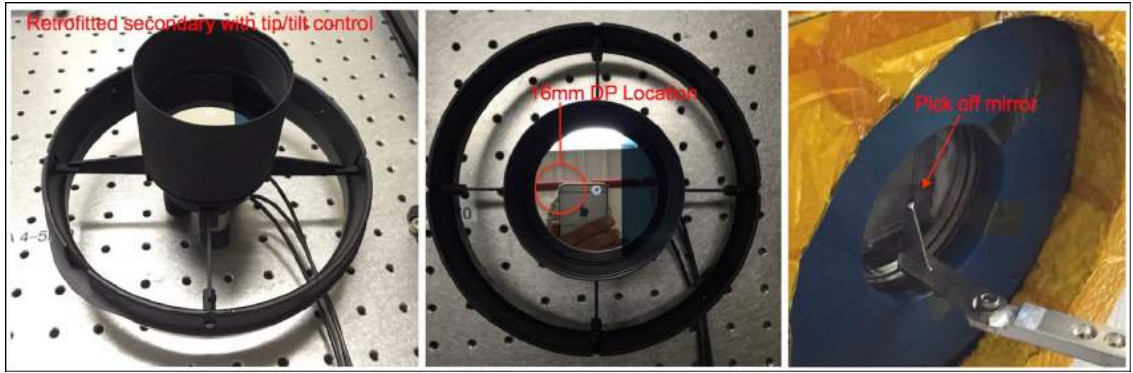
Figure 3.2. The collimator telescope secondary was retrofitted to add tip/tilt control and imprint the DP on its curved surface. The image on the right shows the coronagraph pick-off mirror.

aperture, a 25mm diameter polarizer is placed together with 20mm stop.

The imaging telescope is the core of the *Astrometry instrument* that re-images the source with a 1-to-1 magnification with an f/68.5 beam. The increase in f/# is the result of reducing the telescope aperture. A small pick-off mirror located 80mm upstream of the focal plane directs the light to a DP Low Order Wavefront Sensor (DP LOWFS), which uses a ZWO120MC camera. Tip/tilt measurements of the incoming beam are used to stabilize the image using the active secondary mirror of the collimator telescope for actuation. Table 3.1 summarizes the system parameters.

*The Coronagraph instrument* is fed with a pick-off mirror, shown in Fig 3.2, which is placed 15mm upstream of the focal plane. This mirror has a square 3x3mm footprint and extracts the target star FoV to perform high-contrast imaging. The extracted beam remains with f/68.5 divergence after extraction, until it is collimated and resized to 18mm using a beam expander before being fed to the coronagraph.

## 3.4 The light source simulator

To simulate a star field we created a light source based on a 100$\mu$m thick tungsten 1" diameter disc with a grid of 21x21 laser drilled 5$\mu$m holes. The holes, which are equivalent to point sources for the astrometry camera resolution, simulate background stars. The whole plate is illuminated with a collimated white light laser beam generated by the NKT Photonics SuperK super

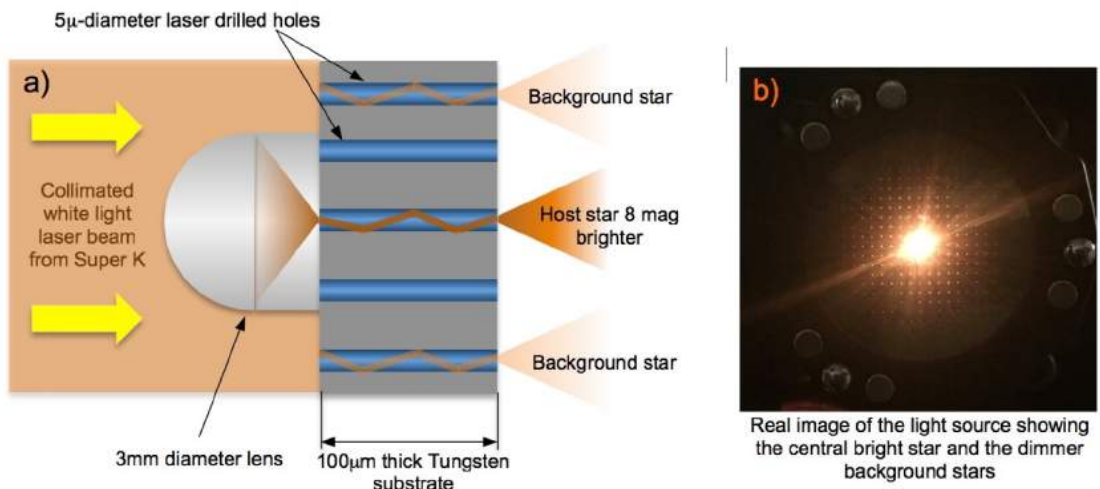
Figure 3.3. a) Schematic of light source construction. b) Image of the light source in operation.

continuum laser operating between 400 and 690nm. On the back of the central hole, a 3mm diameter achromatic lens is glued precisely to place its focus on the center of the central 5$\mu$m hole, creating a simulated star ~1.5x10$^4$ times, or about 10 mag, brighter than the background stars. A schematic of the system is shown in Fig. 3.3. The light source is able to roll and align its rotation axis to simulate telescope roll.

### 3.5 Astrometry module design overview

We have designed the DP to properly sample the astrometry instrument FoV. There are three top-level requirements that control the DP geometry: the instrument FoV, which controls the dot spacing to ensure equally spaced bright spikes over the FoV, the average brightness of background stars, which defines the dot size in order to maintain a similar surface brightness of the spikes and the stars, and the operational wavelength. This experiment has 37.2'x37.2' FFoV, with background stars 10 magnitudes dimmer than the target star and it operates in visible light between 400 to 690nm. The DP geometry on the telescope secondary is shown in Fig. 3.4.

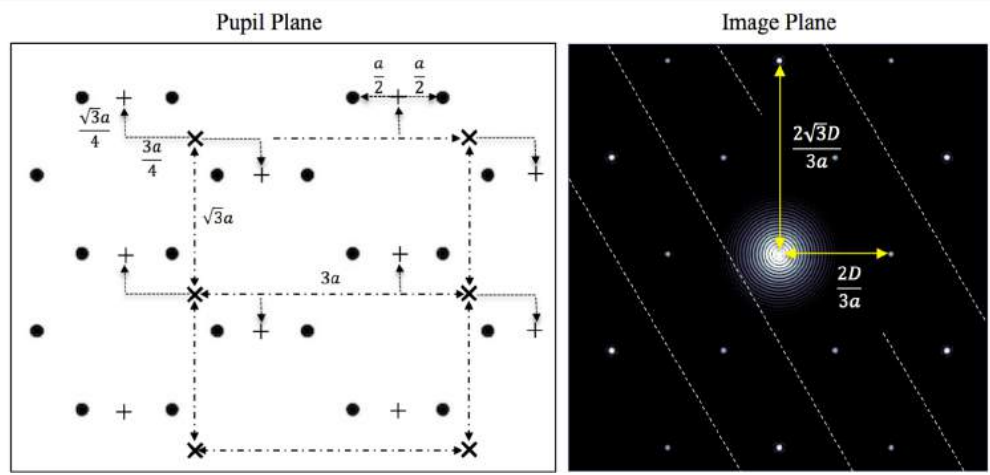

Figure 3.4. On the left, an image of the hexagonal arrangements of dots placed on the pupil is shown. Here, the side of the hexagon is defined as *a*, so the hexagon width is 2*a* wide. On the right, the resulting image and spots spacing is shown. The dashed lines represent zeros of the cosine modulation that eliminates the spot at D/3a.

To calculate the position and size of the dots in the DP we model the hexagonal pattern as a replication of pairs of delta functions over the pupil plane. This can be mathematically represented as,

$$g(x,y) = A\, comb\left(\frac{xD}{3a}, \frac{yD}{\sqrt{3}a}\right) * \left[\delta\delta\left(\frac{xD}{\frac{3a}{4}}, \frac{yD}{\frac{\sqrt{3}a}{4}}\right) * \delta\delta\left(\frac{xD}{a/2}\right)\right], \qquad (1)$$

where the *comb* function is a two-dimensional array of delta functions, $\delta\delta$ represents a pair of delta functions, "*" represents convolution, and *A* represents the scaling factors required to maintain the normalization of delta functions after a Fourier transform. The axis coordinates *x* and *y* have been multiplied by the aperture *D* to normalize the result to the aperture size. At the image plane the Fourier transform of this grid is obtained as,

$$G = F_\xi F_\eta [g(x,y)] = comb\left(\frac{3a}{D}\xi, \frac{\sqrt{3}a}{D}\eta\right)\left[\cos\left(\frac{3\pi a}{2D}\xi + \frac{\sqrt{3}\pi a}{2D}\eta\right)\cos\left(\frac{\pi a}{D}\xi\right)\right], \qquad (2)$$

where $\xi$ and $\eta$ represent the transform variables and axes in the image plane in units of $f\lambda/D$. At the image plane we obtain a bi-dimensional *comb* function with spacing $D/3a$ along the $\xi$ axis. However, this grid is modulated by two cosine functions. The first cosine is bi-dimensional and has a period of 4D/3a along the $\xi$ axis, creating zeros at D/3a and D/a reducing the special frequency of the *comb* by half, allowing deltas only at

$$\xi = \frac{2}{3}\frac{D}{a}, \quad (3)$$

and at integer multiples of this value. Using the same rationale for the $\eta$ axis, we obtain delta functions at

$$\eta = \frac{2\sqrt{3}}{3}\frac{D}{a}. \quad (4)$$

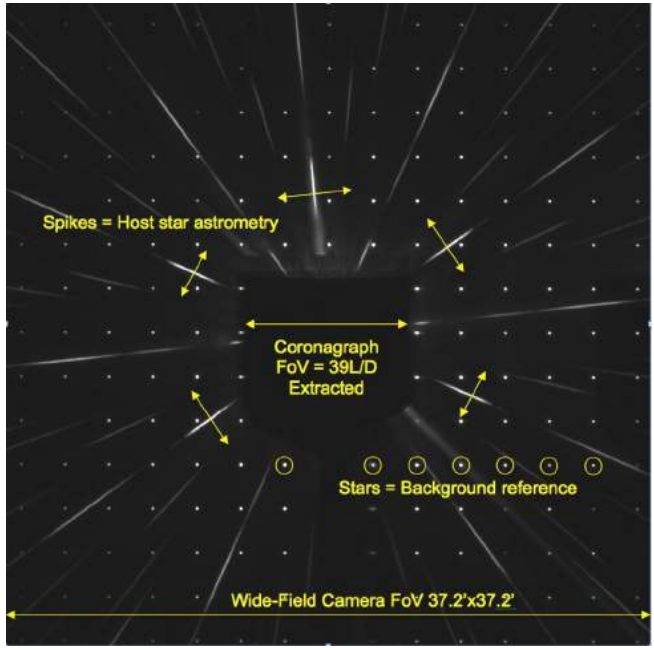

Figure 3.5: Typical astrometry camera image

Similarly, the deltas are replicated at integer multiples of this value. The spot brightness is modulated by the second cosine of Eq. 2, which has a period of *2D/a* but does not change the spatial frequency of the grid because it does not have zeros matching the *comb* period. The geometry described is shown for pupil and image plane in Fig. 3.4.

We designed the system to have spikes starting at 56λ/D and 96λ/D for the first and second order respectively. This could be achieved with a DP composed of 12$\mu$m dots with a spacing $a$=240$\mu$m spacing over a 20mm pupil. However, we needed to place the DP in the first optical surface, which is the secondary of the collimating telescope. As a result, we had to rescale the pupil to its footprint on the secondary, resulting in spacing $a$=120$\mu$m spacing over a 10mm pupil.

Downstream of the secondary mirror, the primary collimates the beam, which feeds the imaging telescope forming an f/68.5 beam that creates an image plane, which is sampled by an APOGEE Alta U16000 camera. This camera has 7.4 $\mu$m pixel size CCD and 16 bits dynamic range. The HFoV is limited to 18.6' on the camera readout by only reading 2000x2000px per frame, which is equivalent to 37.2' or slightly more than half a degree. A typical astrometry image is shown in Fig. 3.5 where the FFoV exhibits diffraction spikes of diminishing brightness as the angular separation from the target star increases. The array of background stars also diminishes in intensity as the FoV increases, but this is caused by residual Gaussian intensity profile of the illuminating beam. The pick-off mirror is distinguishable as a diffuse square shadow in the center that connects with a vertical shadow caused by a 500$\mu$m thick aluminum holder.

### 3.6 Coronagraph design

#### 3.6.1 PIAA lenses and optical layout

The coronagraph is fed with the beam extracted by the pick-off mirror located in front of the astrometry camera focal plane. This beam is expanded to a diameter of 18mm and collimated using a beam expander. The collimated beam is routed to the coronagraph using three flat fold mirrors, of which the last one has active tip/tilt control Piezo actuators to steer the beam's incidence angle entering the coronagraph. The coronagraph's first element is the PIAA apodizer [Guyon 2003] that removes the point spread function (PSF) diffraction rings. The apodizer consists of two lenses separated by 100mm, and a 14.5mm aperture stop placed downstream of the PIAA lenses. The stop removes the light contamination caused by imperfections beyond the lenses' clear aperture of 16mm. The PIAA lens used for the AD were designed for the first experiment at the ACE lab [Belikov et al 2009].

### *3.6.2 Deformable Mirror*

The apodized beam is focused using a 500mm focal length lens placed immediately after the stop and 200mm downstream of the lens there is a BMC Kilo Deformable Mirror (DM). In this plane the beam footprint is 8.7mm in diameter. Hence, it underfills the 32x32 actuator DM, which has a square shape of 9.9 x 9.9mm, resulting in 28 illuminated actuators across the pupil.

A miniaturized DM controller developed at NASA Ames Research Center [Bendek et al 2016] was used to control the DM. On this bench, the DM is directly attached to the controller which is supported by translation and rotation stages, allowing precise beam centering on the DM surface and tip/tilt control to align the beam downstream of the DM. Power and control are supplied using a 12V jack and a USB3 cable for data. The system generates a maximum of 8W of heat, which is removed using a heat exchanger below the controller shown in Fig. 3.6. However, the actuators' stroke was only partially used, reducing the average power consumption to less than 5W most of the time.

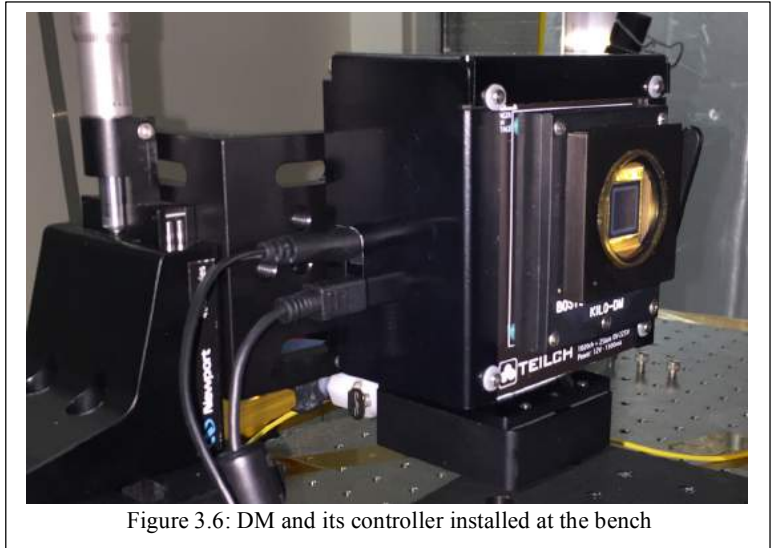

Figure 3.6: DM and its controller installed at the bench

### *3.6.3 Focal plane occulter and LOWFS*

The DM folds the converging beam, which focuses on the Focal Plane Occulter (FPO), located 300mm downstream of the DM. The FPO is made of a fused silica substrate coated with an Optical Density (OD) ~5 chrome reflective mask on top that has the C-Shape shown in Fig. 3.7. The coronagraph's Dark Zone (DZ) is created on the "transparent" region where the planet light can be transmitted and the Wavefront Control (WFC) algorithm can suppress speckles. The inner radius "$r_{in}$" is 98μm, corresponding to 1.8λ/D and the outer radius "$r_{out}$" is located at 363μm or 6.6λ/D. The distance "d" is 45μm and the rotation angle theta is 180°. The inner working angle (IWA) of the system is located at 1.6λ/D, which is where we measured that 50% of the PSF flux reaches the detector. The IWA calibration procedure is described in section 5.4. (Note that the IWA is slightly smaller than $r_{in}$ because the curvature of the IWA edge of the mask is non-negligible compared to the PSF size.) The FPO used for this experiment has an opaque/non-reflective spot in the center of the occulter, limiting the amount of rejected light to the LOWFS to an annular ring. We use a PID control loop that measures the position of the rejected light on the camera detector (ZWO178) to stabilize the pre-PIAA tip/tilt mirror closed loop. We normally run the control loop at 1Hz, allowing us to remove slow drift but not jitter caused by vibrations.

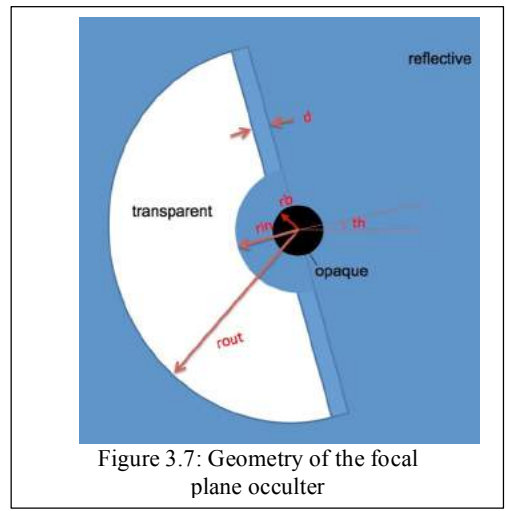

Figure 3.7: Geometry of the focal plane occulter

The light transmitted through the transparent section of the FPO continues toward a 400mm focal length lens that collimates the light and directs it towards a Lyot Stop. The stop suppresses the diffracted light resulting from the PIAA apodization cut-off. Afterwards, the light goes through a second polarizer before reaching the camera (QSI 543 that has 5.4$\mu$m pixels and 16bit dynamic range).

### 3.6.4 Wavefront Control algorithm

We aligned the system using a flat mirror instead of a DM that provided us with the best focal plane PSF not affected by the DM flattening errors. After the optical path and PIAA alignment was completed, we replaced the mirror with a Kilo DM for which we previously created a flat DM voltage map. To establish this map, a procedure was developed for obtaining direct interferometry measurements of the DM shape using the ZYGO interferometer. In this procedure an iterative correction to the applied voltages was calculated based on a simple quadratic model of the DM deflection curve applied to the measured DM shape. Once the mirror shape was measured, a solution was computed, applied, and a new measurement was taken that can be used to calculate the next correction. The procedure provides fast (3-5 iterations) convergence to a stable solution (Fig 3.8) that minimizes wavefront errors by using a constant DM voltage as a starting point. The localized dots visible in Fig 3.8 correspond to actuators working abnormally.

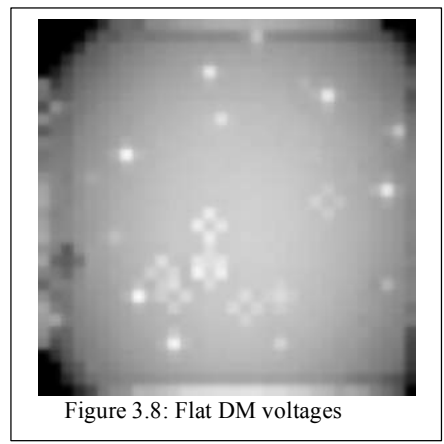

Figure 3.8: Flat DM voltages

Aberrations induced by mid-spatial frequency polishing, alignment, and DM shape errors, as well as residual diffraction, prevents us from achieving the required contrast level when a coronagraph is used without wavefront control. A DM controlled by a Wavefront Control algorithm is necessary to create a dark region around a "star". Two widely used wavefront control techniques include Electric Field Conjugation (EFC) [Give'on et al 2007] and Speckle Nulling (SN) [Trauger et al. 2004] methods. In our experiments the SN wavefront control algorithm was used. Although the SN algorithm is slower in comparison with EFC, it does not require any well-established a-priori optical system model to be used during wavefront correction.

For the iterative SN algorithm, a set of N dark zone brightest speckles is selected in the science camera focal plane to be nulled for each iteration step (Fig. 3.9). Each speckle position in the focal plane can be mapped to the spatial frequency on the DM that produces a speckle in the same location as the nulled speckle but with opposite phase. All the speckles to be nulled are then interfered with a set of sinusoidal DM shapes, which scans in both phase (sinusoidal) and amplitude. The phases and amplitudes of the DM probe that interferes destructively within each of simultaneously corrected speckles are chosen to correct all selected speckles per iteration.

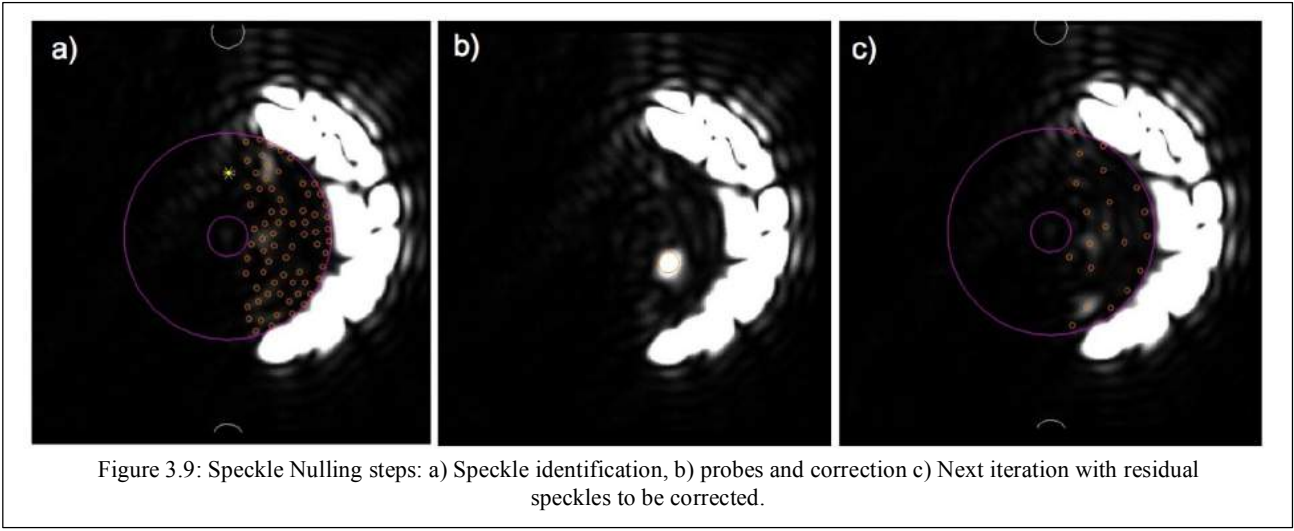
Figure 3.9: Speckle Nulling steps: a) Speckle identification, b) probes and correction c) Next iteration with residual speckles to be corrected.

The speckle position to the DM frequency mapping (6 parameters including the PSF center position) can be established by a simple geometrical calibration procedure that matches a set of calibrated DM frequencies to a set of corresponding speckle positions measured in the science camera focal plane. The long-term PSF drift is measured creating a calibration speckle in the Dark Zone.

The optimal number of speckles to be corrected in the dark zone is on order of $S/(2\lambda/D)^2$, where S is that Dark Zone area. This estimate assumes that the closest focal plane speckles are almost not interacting during any given iteration. However, the algorithm works even in a case when the number of corrected speckles in much larger than the optimal number and speckles influence each other during a wavefront control iteration. Changing N from a few speckles to 2 or 3 times $S/(2\lambda/D)^2$ speckles is a good practice to avoid algorithm stagnation near existing local solutions/minima. SN constraints such as the number of corrected speckles, a radius in which energy is minimized, and the maximal amplitude of the corrective sinusoidal wave set the aggressiveness of the SN algorithm. For example, smaller applied DM strokes and larger separation between speckles reduces correlations/interaction between speckles being simultaneously corrected.

Our strategy was to use more aggressive settings for early iterations. These include a larger number of speckles (2 to 3 times $S/(2\lambda/D)^2$) and higher control loop gain. As a deeper contrast level is attained and the closest speckles start to affect each other, the algorithm aggressiveness was reduced to minimize speckle interaction during probing. When a contrast plateau was reached, the algorithm aggressiveness was switched between three different modes (small number of speckles, optimal number of speckles, and large number of speckles) to avoid algorithm stagnation.

## 3.7 Thermal enclosure and beam stabilization

The astrometry and coronagraph instruments of the AD require high stability to work properly. First, the astrometry experiment requires the image registration for different epochs to stay within one pixel. Larger shifts would affect the performance of the data reduction algorithm presented in the next section. Second, the coronagraph is sensitive to jitter since it causes light leak around the FPO as a function of PSF excursion's amplitude. A typical coronagraph working at $1.6\lambda/D$ requires a PSF stability in the order of $1\times10^{-2}\lambda/D$. In addition, long term drifts cause beam walk on the optics, modifying the speckle field morphology.

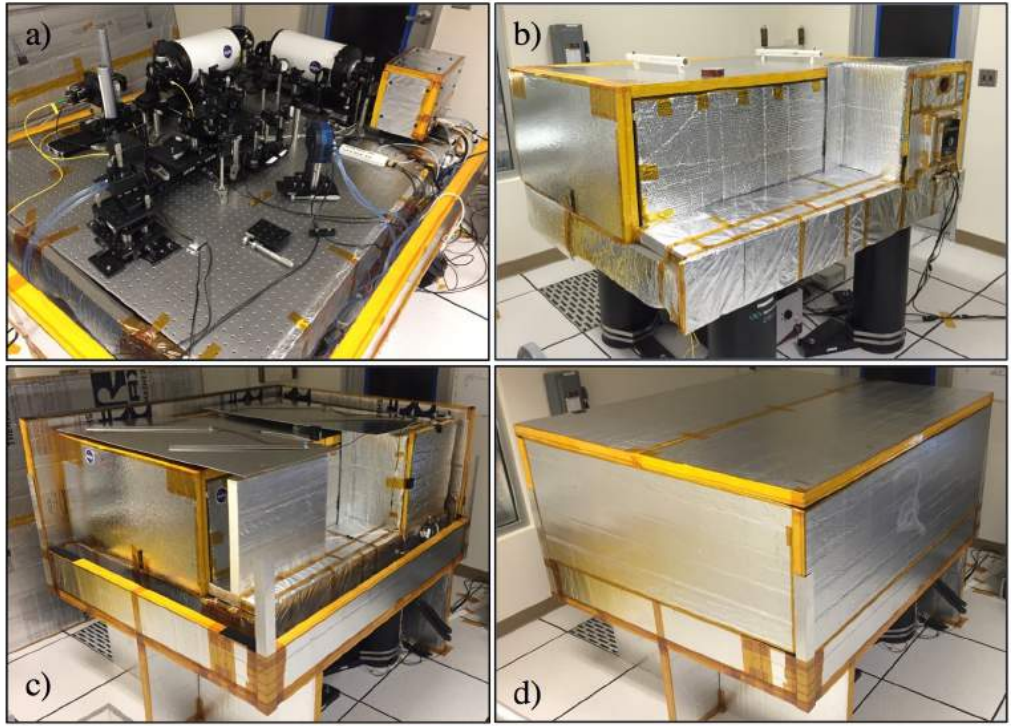

Figure 3.10. Optical bench and lower enclosure frame is shown in frame a). The enclosure consists of a double layer system, where the first layer encloses the entire system, including the optical bench as shown in figure b). Then, we mounted thermal heaters in an air gap between the inner enclosure and a second enclosure around the first one as shown in figures c) and d)

The Air Conditioning (AC) bang-bang temperature controller causes temperature oscillations in the lab of 0.25°C PV every 15 minutes (Fig. 3.11). Those temperature changes induced a drift of up to 7 pixels on the astrometry camera, created turbulence, vibration, and acoustic effects that were unmanageable for the WFC. These environmental conditions required the implementation of thermal control and turbulence shielding in addition to using active beam stabilization. We decided to build the enclosure shown in Figure 3.10, which has two nested enclosures with a thermal control system regulating the air temperature in the cavity between them. External thermal loads are shielded and absorbed by this layer of air. We also installed water-cooling to remove the heat of internal sources such as the astrometry camera, the coronagraph camera and the deformable mirror controller.

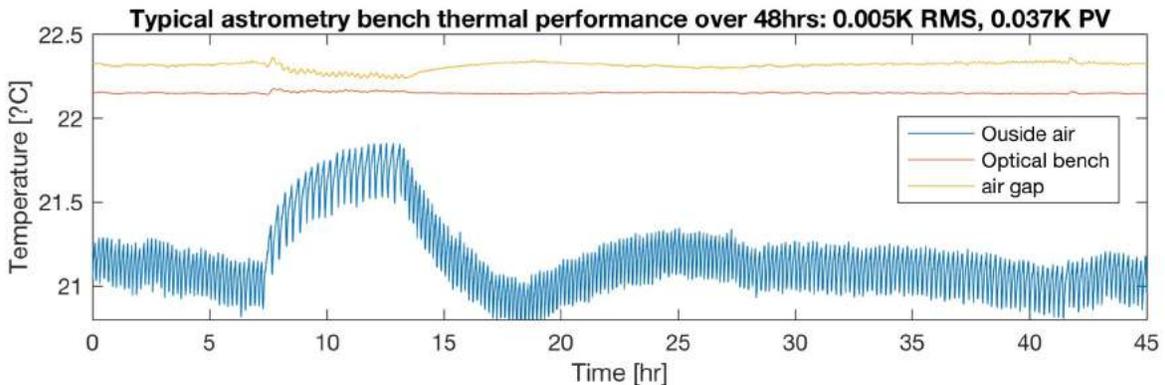

Figure 3.11. Temperature evolution for the clean room air called "outside air" in the legend, the optical bench, and the air gap during a typical 48hrs run.

The bench is hot biased, i.e. the optical elements and the breadboard are maintained about 1°C above the ambient temperature. We implemented a PID control loop in LabVIEW, which uses the heater plate temperature between the enclosures as the control input signal. After system is in steady state it is possible to improve the performance by changing the input signal to the breadboard temperature. During steady-state operation the control loop supplies between 5 to 10W to maintain the hot-biased bench with enough control authority to absorb room temperature changes.

Fig 3.11 shows the temperature evolution of the room temperature or "outside air" in blue, which has a 0.25K saw tooth as a result of AC bang-bang controller operation. Also we observe larger oscillations that probably respond to some kind of energy save mode for the whole building. The red line shows the optical breadboard temperature, which has 5mK RMS and 37mK PV over 48hrs. Similar performance was observed for every run.

# 4 Astrometry data reduction

## 4.1 Data reduction algorithm

The data reduction algorithm only measures the angular component of the spike positions because radial smearing of the diffractive features that create the spikes prevents accurate radial measurements. Nevertheless, the Cartesian X-Y astrometry vector can be obtained by projecting multiple angular displacements into X-Y coordinates.

The algorithm inputs are images from two different epochs, $I_1$ and $I_2$. First, the algorithm calculates a reference image, $I_{ref}$, as the sum of $I_1$ and $I_2$, and a difference image, $I_{diff}$. The next step in the data reduction process is to calculate the angular derivative of $I_{ref}$, which provides the relationship between small image rotations and corresponding change to $I_{ref}$.

To obtain the angular derivative, we first compute the Cartesian unitary derivatives of the reference image, $I_{ref}$. For a pixelated image, we employ discrete differential operators that approximate image derivatives. There are many choices for these operators, which vary in accuracy, direction preference, and functionality in the presence of noise, among others. We seek to increase the derivative sensitivity while providing some crude noise smoothing by convolving $I_{ref}$ with the 3-by-3 isotropic Sobel operator. The x-derivative is shown as follows

$$\frac{\partial I_{ref}}{\partial x} = \alpha(I_{ref} * f_x), \tag{5}$$

where α is a calibration factor and $f_x$ is the normalized Sobel operator in the x-direction shown in MATLAB matrix format

$$f_x = \frac{1}{8}[-1\ 0\ 1; -2\ 0\ 2; -1\ 0\ 1]. \tag{6}$$

The Cartesian derivative terms are used to compute the angular derivative

$$\frac{\partial I_{ref}}{\partial \theta} = -y\frac{\partial I_{ref}}{\partial x} + x\frac{\partial I_{ref}}{\partial y}. \tag{7}$$

Then, the angular displacement is computed by dividing, pixel-by-pixel, the difference image, $I_{diff}$, by the angular derivative, given by Equation (8):

$$\theta_{disp} = \frac{I_{diff}}{\frac{\partial I_{ref}}{\partial \theta}}. \tag{8}$$

At this step, even with some smoothing provided by the Sobel operator, the noise on image areas where no spikes are present is orders of magnitude higher than the signal to be measured, which is in the range of $10^{-2}$ to $10^{-5}$ pixels. To solve this problem, the Signal to Noise Ratio (SNR) of the angular distortion measurement is computed for every pixel. The signal is the value of the angular derivative and to first-order the noise is the root sum square of the Read Out Noise (RON) plus the photon noise,

$$SNR = \frac{\frac{\partial I_{ref}}{\partial \theta}}{Noise}. \tag{9}$$

The angular distortion image is multiplied with the SNR squared, amplifying the values along the spikes and minimizing them on the background,

$$\theta_{disp\_SNR} = \theta_{disp} SNR^2. \tag{10}$$

The image containing the angular distortion is noisy, especially along the spikes. A binned version is created to reduce the noise level and the computational power required to process them. This process does not discard useful information because the spikes remain resolved. Then, the angular distortion image is divided by the binned version of $SNR^2$ to recover the correct values on the angular distortion image. The pixel value represents the angular distortion for its location in units of pixel size, i.e. a value of 1 represents 7.4µm of angular distortion at that detector location.

The binned angular distortion obtained only contains information available on the spikes location. The values for the pixels between the spikes are obtained using a kernel convolution interpolation. The $SNR^2$ and the $\theta_{dist\_SNR}$ are binned by a factor of ten to obtain the $SNR^2_{bin}$ and images of $\theta_{dist\_SNR\_bin}$ (250x250 pixels), which can be interpolated by performing the convolution with a function $g$

$$\theta_{dist\_SNR\_bin\_interp} = (\theta_{dist\_SNR\_bin}) * g, \tag{11}$$

where $g$ is a Gaussian kernel defined as

$$g = e^{-\left(\frac{x^2+y^2}{2\sigma^2}\right)}. \tag{12}$$

In Equation 12, $\sigma$ defines the FWHM of the Gaussian kernel. Controlling $\sigma$ will define how aggressive the interpolation is and therefore $\sigma$ is set as a parameter in the algorithm that can be adjusted to match the highest spatial frequency distortion expected in the system. Finally, to recover the real values of the angular distortion, it is necessary to divide the $\theta_{dist\_SNR\_bin}$ by the $SNR^2_{bin\_interp}$ matrix,

$$\theta_{dist\_real} = \frac{bin(\theta_{dist}SNR^2)*g}{bin(SNR^2)*g}. \tag{13}$$

The kernel size is selected to perform the best interpolation between bright spikes at large field angles where most background stars will be found. As we move closer to the center field, the distance between the bright spikes is reduced linearly with the field angle. As a result, the interpolation kernel becomes too large for small fields in the image inducing interpolation errors. To solve this problem, a spatially varying kernel size *g* can be used to reduce its size, as the interpolation gets closer to the center of the image.

**4.2   Algorithm validation**

The algorithm described in the previous section was validated using real data. The test consisted of copying the first epoch real image and shifting it by 0.5px to the right. Hence, the real distortion to be measured is known, and the algorithm can be evaluated.

Since the spikes do not uniformly sample the distortion and there is no data inside the circle blocked by the occulter, the interpolation result might be biased as a consequence of changing the Gaussian kernel size. If the kernel is too large, it will average valuable high-spatial frequency distortions. A kernel size with FWHM = 55 was set to provide Nyquist sampling of the spikes in the outer part of the image where most of the stars can be found.

To characterize the effect of the kernel size on the resulting distortion, the algorithm was applied to the spikes and the stars independently for different kernel sizes on a real image and its 0.5px-shifted version. Since there is no real astrometry signal, just an image displacement that is common to both features, the result should be the same for the spikes and the stars, which are perfectly stable in this case.

**4.3   Algorithm optimization**

Using the parameters from the previous section alongside the milestone data, the algorithm is characterized by creating a "known distortion" data set. The data set is constructed by taking the first non-dark image of the milestone data as the first epoch in a series of artificially induced sub-pixel shifted epochs. Each subsequent epoch is created by shifting the first epoch by a known amount using an interpolation method. (The resulting image sequence represents uniform known translational motion, except that the noise in all the images is also translated, while in real life it would be randomly drawn.) For purposes of this discussion, we define *"relative"* measurement as a cumulative sum of consecutive measurements, equally spaced, and an *"absolute"* measurement as the difference between the first and the last epoch of the same data sequence.

The interpolation method employed to create the "known distortion" data set makes use of the Fourier Shift Theorem; the accuracy of this method was verified to the order of $1 \times 10^{-13}$ mean separation from the desired sub-pixel shift for shifts as small as 0.25% of a pixel. To perform this test we compared the image obtained after "n" shifts that results in one full pixel shift, with move the pixels columns (or rows) to the next one. For example, to validate 0.25% pixel steps we shifted the image 400 times.

Next, the algorithm is executed on this data set in three separate configurations: first, using the simple derivative approximation of subtracting a one-pixel shifted version of $I_{ref}$ from itself with respect to X and then to Y; second, using convolution with the Sobel operator to compute the derivative; and third, using the Sobel derivative approximation with a calculated calibration factor.

Using the simple derivative approximation in the algorithm, a nonlinear response results from the computed relative and absolute spike distortions in tip and tilt. This response demonstrates that the sensitivity to small shifts by the simple derivative approximation is insufficient to obtain accurate tip/tilt estimates. The calculation of the angular derivative and the $SNR^2$ map are dominated by noise, biasing the interpolation stage of the algorithm and its results. After testing, it was determined that this nonlinear response is most evident for pixel shifts on the order of 1% of a pixel; the "known distortion" data set used is constructed to demonstrate this effect. The shift between any adjacent images in the set is 0.667% of a pixel to the right, and 0.5% of a pixel down, with a total shift of 0.1333 pixels to the right and 0.1 pixels down from the first image to the last.

Fig. 4.1a displays the results of the algorithm on the "known distortion" data set. The relative tip/tilt lines are both curved. The absolute tip/tilt lines show a nonlinear behavior for the first few images, and then become linear.

This is expected because the absolute distortion is computed with respect to the first image, so after three images, the total shift is larger than 1% of a pixel. The relative distortion is computed from image to image, so the shift stays below 1% of a pixel, making it more susceptible to nonlinear effects.

Fig. 4.1b shows the results of the algorithm using the Sobel operator to approximate the derivative. The relative and absolute spike distortions for tilt are coincident; meanwhile the

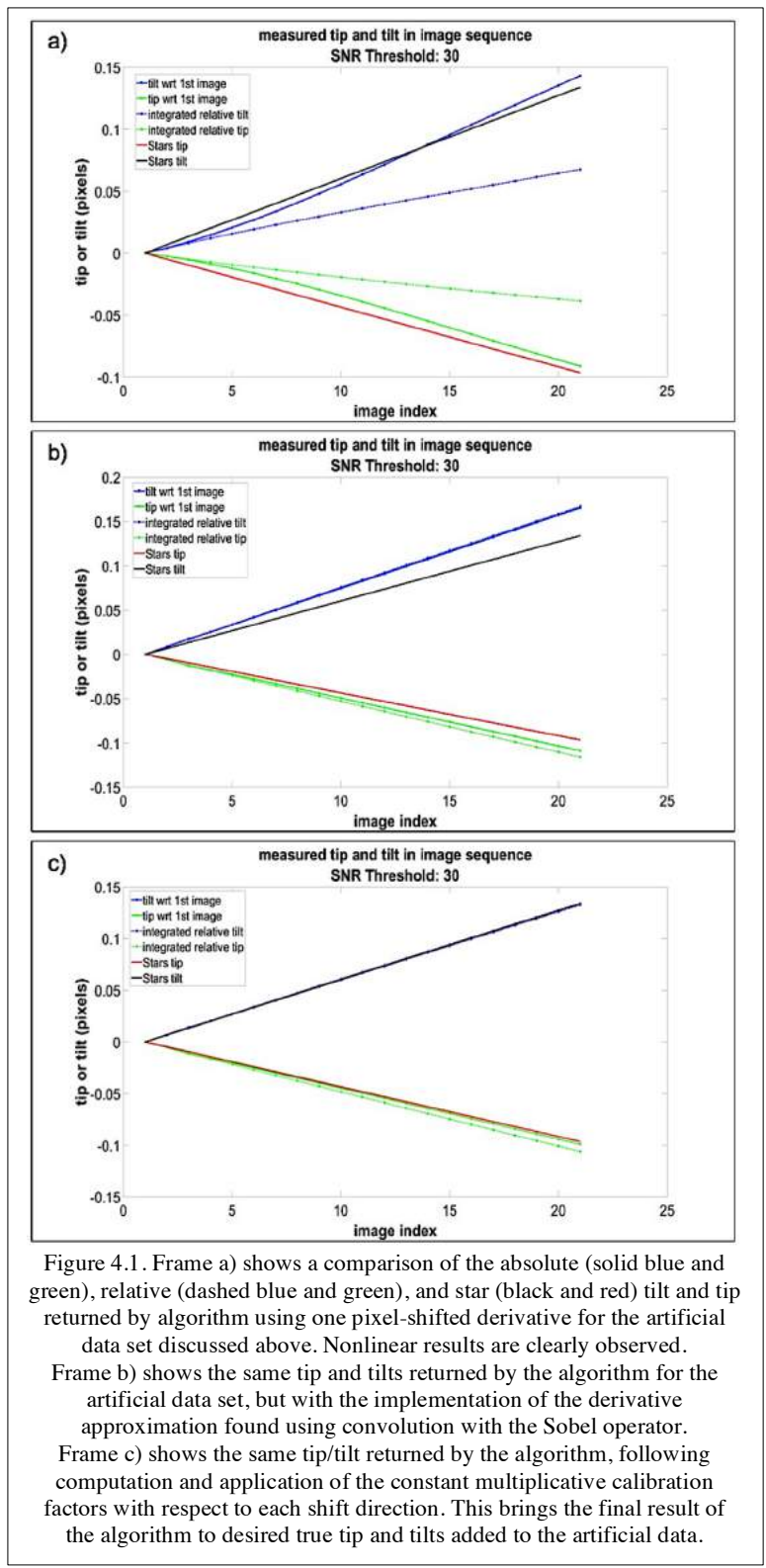

Figure 4.1. Frame a) shows a comparison of the absolute (solid blue and green), relative (dashed blue and green), and star (black and red) tilt and tip returned by algorithm using one pixel-shifted derivative for the artificial data set discussed above. Nonlinear results are clearly observed. Frame b) shows the same tip and tilts returned by the algorithm for the artificial data set, but with the implementation of the derivative approximation found using convolution with the Sobel operator. Frame c) shows the same tip/tilt returned by the algorithm, following computation and application of the constant multiplicative calibration factors with respect to each shift direction. This brings the final result of the algorithm to desired true tip and tilts added to the artificial data.

spike distortion lines for tip trend linearly after the first few images. This is a demonstration of the improved sensitivity provided by the Sobel operator derivative approximation with respect to small pixel shifts compared to the previous method. The nonlinear effects are greatly minimized, having been nearly removed entirely in the tilt, and only have a nonlinear "bump" at Image Index 3 in the tip before stabilizing to a nearly constant slope.

To make further improvements between the star centroid tip/tilt and spike distortion estimates, calibration factors are computed separately for the star centroids as well as the Cartesian unitary derivatives. The slopes between adjacent epochs along each star and spike curve are computed and the results are averaged together. The computed average functions are included as a multiplicative calibration factor to correct for non-optimal estimation of the star or spike distortions. For the star centroid tip/tilt, the calibration factor is applied to the computed shifts as the last step of the algorithm. This removes any significant bias appearing in Fig. 4.1c (bottom) due to the centroiding algorithm. For the derivatives, the calibration factor is applied to Eq. 10. The calibration factor for the derivatives is computed iteratively by repeated application of the algorithm until multiplication of successive calibration factors from each iteration converge to the known distortions beyond a predetermined threshold level. When this has occurred, the respective spike distortion curves become nearly coincident. The results of this process are seen in Fig. 4.1c. It should be noted that the calibration factor must be recomputed if changes are made to the algorithm, the geometry of the testbed, the binning size, the kernel size, or the SNR threshold. However, the algorithm is robust when utilized for astrometry measurements because no significant change on the telescope or the objects is expected between epochs.

Further optimization of the algorithm is being considered. This includes masking the stars/spikes to calculate their distortion estimates separately. If done carefully, this should help to alleviate the need for the multiplicative calibration factors, as the presence of the stars in the spike calculations provides a portion of the bias that the calibration factor is removing. Further optimization can include creating an adaptively sized interpolation kernel that can have a time vary in time from epoch to epoch, and to what size it should change to while interpolating.

# 5 Milestones

The experiments and measurements performed using this testbed are designed to meet the TDEM milestones. Completion of these milestones is documented in this report and reviewed by the Exoplanet Exploration Program.

## 5.1 Milestones definition

This project has two milestones, which read as follows:

**Milestone #1 definition:**

**Broadband medium fidelity imaging astrometry demonstration**

*Demonstrate $2.4 \times 10^{-4}$ $\lambda/D$ astrometric accuracy per axis performing a null result test. The laboratory work will be carried out in broadband spectrum covering wavelengths from 450 to 650nm using an aperture pupil (D) equal or larger than 16mm.*

The "angular separations" are defined in terms of the source wavelength $\lambda$, and the diameter $D$ of the aperture on the DP, which is the pupil-defining element of the imaging astrometry camera. For this milestone, a DP simulates the telescope primary and pupil and will be illuminated by an array of broadband point sources forming f/25 to f/50 beams with respect to the pupil.

**Milestone #2 definition:**

**Broadband medium fidelity simultaneous imaging astrometry and high-contrast imaging**

*Demonstration of milestone #1, and performing high-contrast imaging achieving $5 \times 10^{-7}$ raw contrast between 1.6 and $6\lambda/D$ by a single instrument, which shares the optical path, from the source to the coronagraphic and astrometry FoV separation. The ability of achieving $5 \times 10^{-7}$ raw contrast will be considered as proof of no contamination of the inner working angle (IWA).*

## 5.2 Astrometry Results

### 5.2.1 System verification

The astrometry experiment was performed first, however the experiments were repeated after the coronagraph arm was completed. The addition of the pick-off mirror impacted the data reduction algorithm, because some spikes were partially vignetted, biasing the amplitude of the motions and preventing measuring small-scale distortions on those locations. In addition, the mirror edges diffract light in different radial directions, resembling diffraction spikes. Careful selection of the spikes SNR threshold was necessary to avoid those sources of noise. There was also an impact for some stars that were partially vignetted by the pick-off mirror spider, rapidly magnifying relative motions of the source with respect to the pick-off mirror. The algorithms were calibrated as described in section 4.3.

### 5.2.2 Configuration during data acquisition

Before every run, the bench was thermally stabilized for at least 24hrs after the enclosure was closed and the thermal PID control loop was started. Also, the cameras and internal sources of heat needed to be turned ON at the same time the thermal control loop was started, avoiding changing thermal loads after the bench reached the temperature set point.

Once the bench was stabilized within 20mK of the set point the astrometry, LOWFS PID control loop was turned ON to stabilize the image within the same pixel for the duration of the run to avoid angular derivative nonlinearities. During a typical run the LOWFS would maintain the PSF stability around 0.7µm (0.2px) PV and ~0.1µm (0.03px) RMS over the complete duration of the run. Fig. 5.1 shows a typical run data where residual PV temperature oscillations of 25mK (black line) will result in secondary mirror tip/tilt Piezo commands of approximately 3 to 4 volts, which translate in angular motions at a rate of 0.68"/V. Data from data set 1 is shown on the left image of Fig. 5.1. The image on the right shows the PSF position in both axes during the run demonstrating the stability of the system down to 0.03px RMS. However, two non-common path elements, the pick-off mirror, and the camera itself, degraded the PSF stability, causing up to 0.3px PV motions on the astrometry camera (7.4µm px), which is ten times larger than the motion in the LOWFS camera, but it is comfortably within the operational regime for the astrometry post processing algorithm.

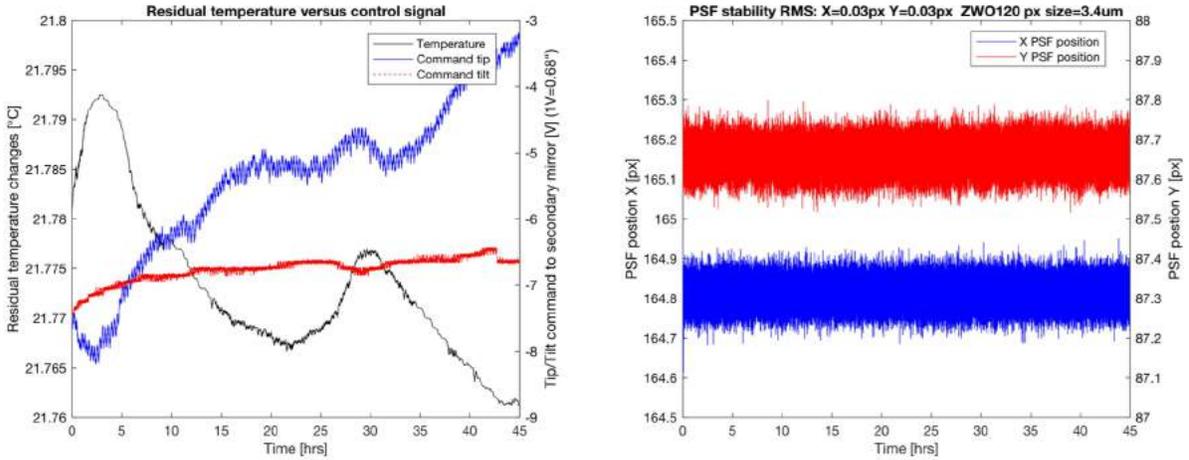

Figure 5.1: Typical astrometry bench thermal and stability performance.

### 5.2.3 Null Test definition

The *Null Test* goal is to determine how accurately the system is able to measure a null astrometry signal where no motion occurs. The residual error of the null test quantifies the ability of the system to perform relative measurements necessary to detect and characterize planets using imaging stellar astrometry at different epochs in time.

Our null test success criterion states that the difference between the spikes and stars tip/tilt, under the condition (or at least assumption) of null displacement of the central star with respect to the background stars, must be equal or smaller than *$2.4 \times 10^{-4}$ $\lambda/D$ RMS*. To successfully meet this milestone #1, the null test should include at least 10 epochs equally spaced in time over 40 hours.

We performed three runs, each 48hrs long, with 12 epochs, one every 4hrs, and used five images per epoch. The bench was stabilized to 22.5°C+/- 20mK over 48hrs. The LOWFS stabilized the PSF to a fraction of a pixel. For example, during run #2 we achieved 0.14px and 0.11px RMS stability on the ZWO camera that has 3.4$\mu$m pixel size, which is equivalent to $1.06 \times 10^{-2} \lambda/D$ and $8.34 \times 10^{-3} \lambda/D$ stability for the X and Y axes, respectively, in the astrometry camera. Maintaining the PSF position stable requires, as mentioned before, motion of the collimating telescope. Those corrections maintain the PSF stable, but cause beam walk, altering the system distortion and thus the star positions. Fig. 5.2 shows the evolution of data set # 2. The image on the left shows in red the PSF motion in X and Y on the astrometry camera detector. The image on the right shows the X and Y projection as function of time. The PV excursions of the PSF can be estimated from both plots to be on the order of 0.3px over 48 hrs. This residual motion is likely dominated by non-common path errors.

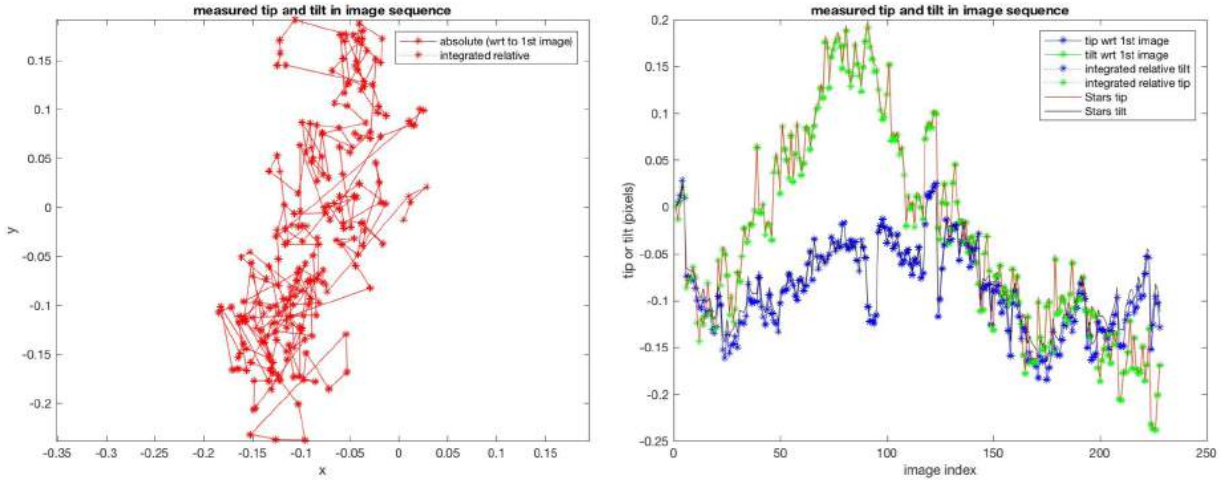

Figure 5.2: The beam walk on the detector plane is shown on the left in red. The X and Y-axis beam walk projection versus time is shown on the right. The blue and green line represent the measurement obtained from the spikes and the red and black is from the stars. At the scale of this plot of 0.45px top to bottom the difference between stars and spikes measurements is indistinguishable.

The plot on the right of Fig. 5.2 also shows and overlaps of the stars only measurements (black and red), and spikes only measurements (green and blue). They seem to overlap given the large 0.45px top to bottom scale of this plot. However, distortion changes create differences on the order of 15 milli px which is imperceptible on Fig 5.2, but easily seen when the spikes and stars data is subtracted as shown on the right of Fig 5.6.

### 5.2.4 Astrometry Null Test

The *Astrometry Null Test* is defined as the relative measurement of the target star with respect to the background stars, when there is no actual motion between the two. This is the measurement relevant for the milestone and the one that would allow us to detect exoplanets. To obtain a stellar astrometry epoch we measure the positions of the background stars and the position of the target star spikes both with respect to the detector.

Given measurements with no noise, the residual would be exactly zero at all epochs, showing the star and spike centroiding are providing equal results. We recover a distortion map of this real residual, and use it to remove the effect in post-processing.

The data reduction algorithm discussed in Section 4 is modified to mask off stars that the recoverable map of the absolute distortion in the focal plane cannot reach (outside a radius of 850 pixels). The algorithm thus provides a map of the absolute distortion, accurate within a radius of 850 pixels, the absolute distortion coefficients (including global average tip/tilt) computed by averaging over this map, and the measured locations of the background star centroids for each epoch. To isolate the distortion, we take small regions in the absolute azimuthal distortion map (Fig. 5.3a) centered at the location of the individual stars, shown in Figure 5.3a, and perform a Zernike fit with Z2 and Z3 (tip and tilt) at each region to estimate the high order distortion in the region.

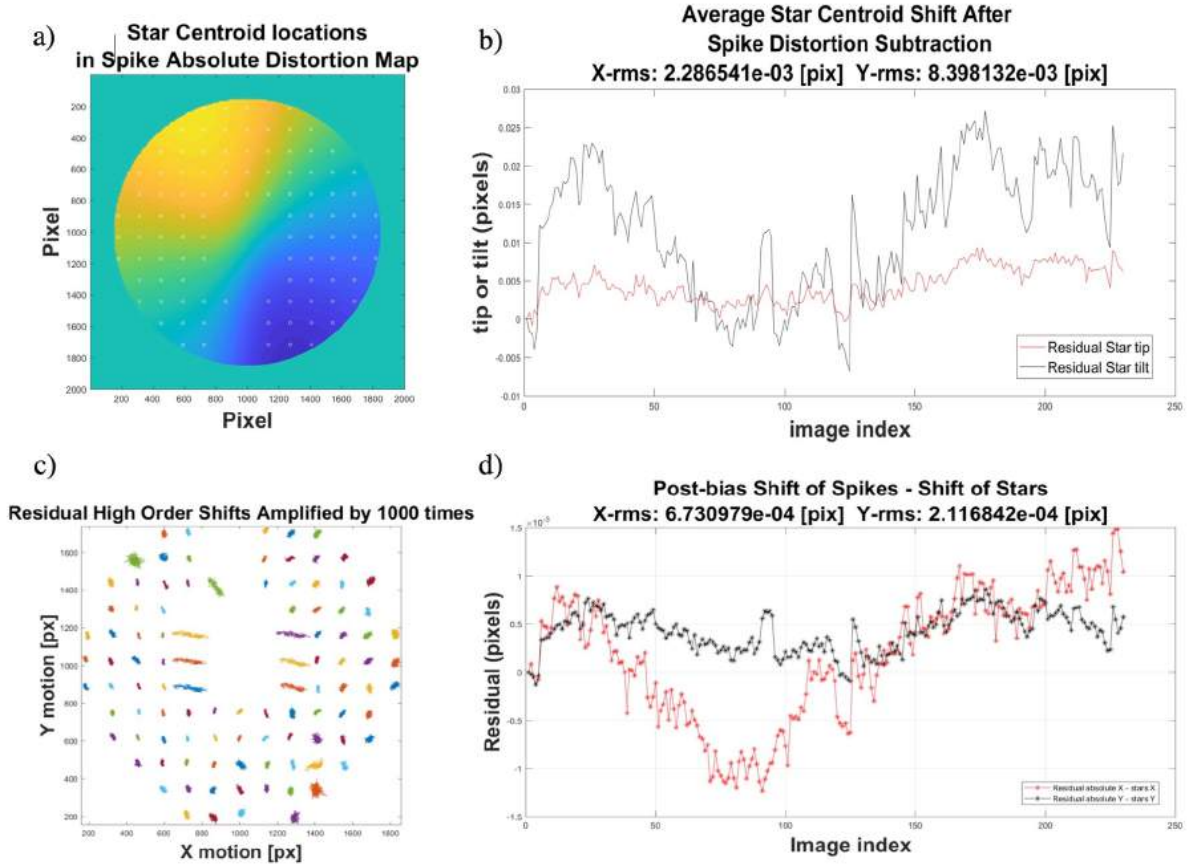

Figure 5.3: a) Distortion map and sampling points. b) Residual error between stars and spikes after performing only tip tilt correction from the spikes. c) High-order term distortion correction residuals. d) Null test result using the tip/tilt and high order distortion correction, which are more than 10 times better than only considering the spikes tip/tilt information and ignoring the high order terms. Note that there are 6.1 pixels per λ/D.

The size of the region is chosen such that the Zernike fit with only two orders sufficiently approximates the behavior of the region. This does not have to be too small because the absolute distortion map does not exhibit rapidly changing behavior. The results of these Zernike fits are then subtracted from their corresponding background star's respective X and Y centroid location, along with the respective global average tip/tilt. This will isolate the residual difference in the centroid location due to beam walk distortion for each individual background star. The plot in Fig. 5.3b shows the average of all the centroids per epoch after this subtraction, and the plot in Fig. 5.3c shows a map of the residual shift for each individual star, per epoch, amplified by 1000 for visual clarity.

This map shows a large distortion for the background stars located near the occulter around the center of the image. This is due to the occulter edge diffracting light and stretching those stars. The unamplified versions of the values shown in this figure are then used as a known bias for the data reduction algorithm to minimize the effect of the distortion. Fig. 5.3d plots the resulting difference between the measurement of the shift of the background stars and the measurement of the shift of the spikes in each direction, and the RMS error for each milestone data set is shown in the rightmost columns of Table 5.1. ***The factor of more than 10 improvement between Fig 5.3b and Fig 5.3d is the result of including the high-order distortion terms in the calibration.***

Table 5.1: Milestone #1, null test data sets summary

|  | Milestone req. ($\lambda/D$) | ($\lambda/D$) Stars w/r to spikes full correction | | Equivalent astrometry accuracy for [uas] | |
| --- | --- | --- | --- | --- | --- |
|  | 1-axis | X-axis | Y-axis | D=2.4m | D=4.0m |
| Set #1 |  | **8.93x10$^{-5}$** | **5.45x10$^{-5}$** | **3.09** | **1.85** |
| Set #2 | 2.4x10$^{-4}$ | **9.62x10$^{-5}$** | **3.02x10$^{-5}$** | **2.72** | **1.63** |
| Set #3 |  | **4.07x10$^{-5}$** | **3.43x10$^{-5}$** | **1.61** | **0.96** |

Note that milestone #1 data set 3 had a slightly different geometry, so the calibration factors used were not as accurate. This leads to the numbers for the difference between the relative and absolute distortions.

### 5.2.5 *Spikes-only registration test*

In order to better characterize key noise terms in our technique, we performed two more tests: First, the *Spikes only registration test*, which allows determining the accuracy with which we can register the position of the spikes with respect to themselves on the detector between different images. Second, the *Stars-only registration test,* which provides information about the centroiding accuracy ***without*** considering the spikes and distortion calibration, so it is the limiting factor for current telescopes. All the tests used the same data sets.

Since we are measuring stars and spikes separately it is not possible to have a reference for the measurement. Therefore, we defined a test that compares the cumulative sum of consecutive measurements, equally spaced, and an absolute measurement between two astrometry epochs as shown in Fig. 5.4. The error is the difference between the absolute, 2-point measurement, and the cumulative sum of *n* incremental measurements between those points.

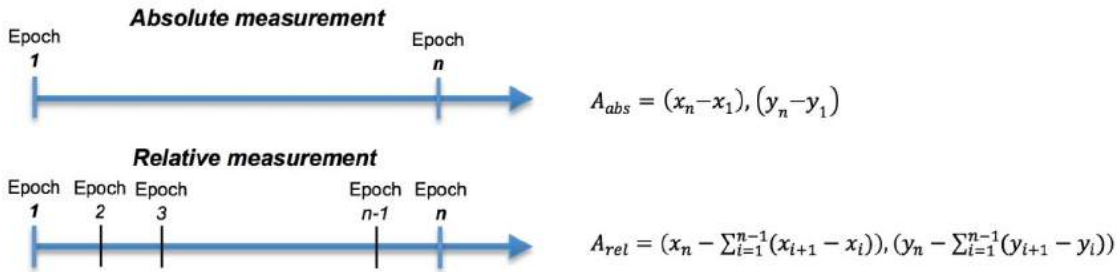

Figure 5.4: Relative vs. absolute measurement

For this test we reduced the data set only measuring the motion of the spikes in X and Y. The results are shown in Table 5.2. We were able to achieve 2.38x10$^{-5}\lambda/D$. Fig. 5.5 shows the residual error between cumulative and absolute measurements used to compute the values in Table 5.2. This result set the accuracy limit, per epoch, that we could expect for this optical bench and camera. Note that these values represent only the spikes contribution to the error budget. To complete the astrometry measurement it is necessary to measure the position of the stars as well.

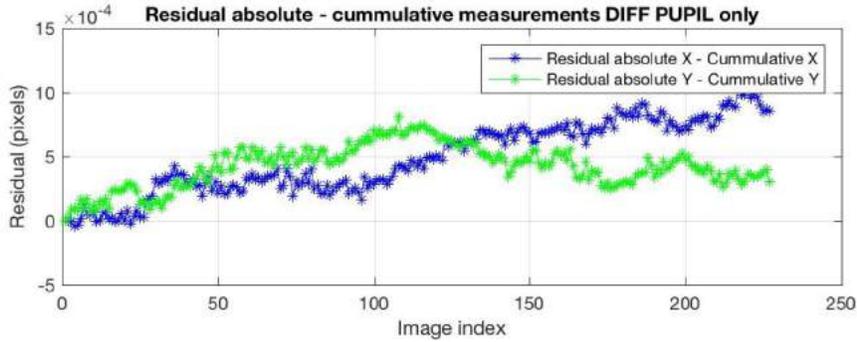

Figure 5.5: Residual error between the cumulative and absolute measurements for each image.

### 5.2.6 Stars-only registration test

The registration test using only the stars shows an accuracy of $7.09 \times 10^{-4}$ λ/D, which is equivalent to 23.4$\mu$as for a 2.4m diameter telescope. The results are shown on the rightmost columns on Table 5.2. The RMS error of the registration test is more than 20 times larger than using the spikes. This result is consistent with the best performance reported by Hubble using precision scanning techniques [Riess et al. 2014]. Also, if we extrapolate this to a 4m telescope such as HABEX, the spikes-only registration test would achieve sub-1$\mu$as accuracy, whereas the stars-only registration test would scale to ~14$\mu$as accuracy, insufficient to find and characterize earth-like planets.

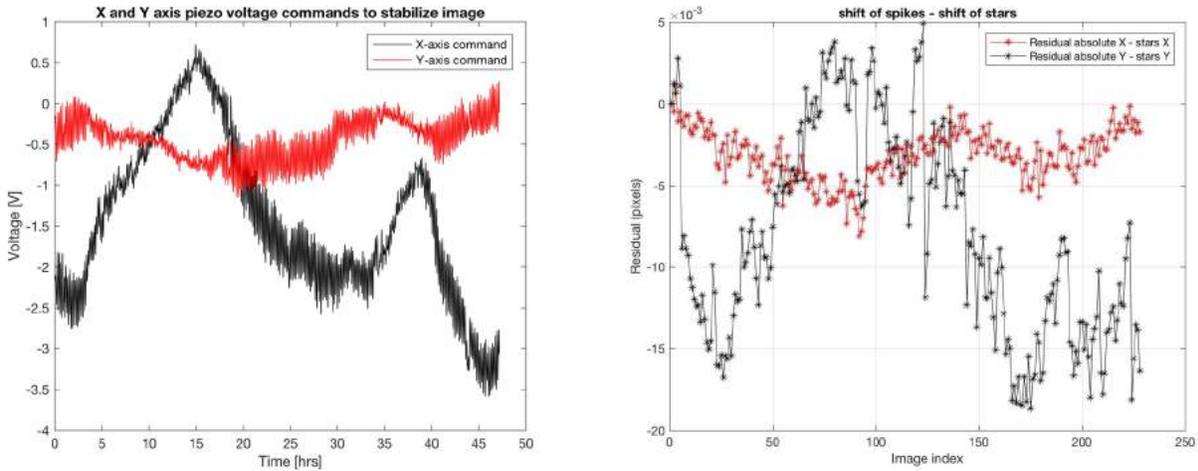

Figure 5.6: Comparison of the control signal applied to the collimating telescope tip/tilt mirror (left) and the difference between the stars' and spikes' position (right). The two plots cover the same period of time. The correlation between tip/tilt and error suggests that it is caused by beam walk that biases the stars' position.

The difference between stars and spikes measurements is caused mostly by distortion changes resulting from beam walk and optics deformations. Fig. 5.6 (left) shows the tip/tilt voltage commands that were applied to the collimator telescope secondary during the run, and Fig. 5.6 (right) shows the difference between the positions measured by spikes and the stars. The correlation is noticeable by inspection. The spikes-only registration test data presented in section 5.2.5 was obtained from the same images and therefore suffered from distortion, however the spikes-only registration test data (Fig 5.5) does not show any correlation with the tip/tilt commands demonstrating the resilience of the spikes to the beam walk perturbations.

| Table 5.2: Spikes-only and stars-only null tests results | | | | | |
|---|---|---|---|---|---|
|  | Spikes-only Null test RMS ($\lambda/D$) | | Milestone #1 ($\lambda/D$) | Stars-only null test ($\lambda/D$) | |
|  | X-axis | Y-axis | 1-axis | X-axis | Y-axis |
| TDEM result | 2.38e-05 | 4.00e-5 | 2.4e-4 | 7.09e-04 | 1.65e-03 |
| Equivalent for D=2.4m | 1.0µas | 1.7µas | 10.0µas | >23.4µas | >39.0µas |
| Equivalent for D=4.0m | 0.6µas | 1.0µas | 6.2µas | >5.79µas | >9.65µas |

### 5.2.7 Distortion maps

Once the tip tilt distortion component is computed and taken into account the high order terms are sampled at each background star location. The distortion map shown on Fig. 5.7 represents the equivalent distortion applied to each star to recover their correct position on the sky. The title of the image displays the amplitude of the largest distortion arrow in the figure. The two distortion maps shown in Fig 5.7 corresponds to two epochs arbitrarily selected to show cases with small (0.004px) and large (0.008px) high-order distortion during the same run.

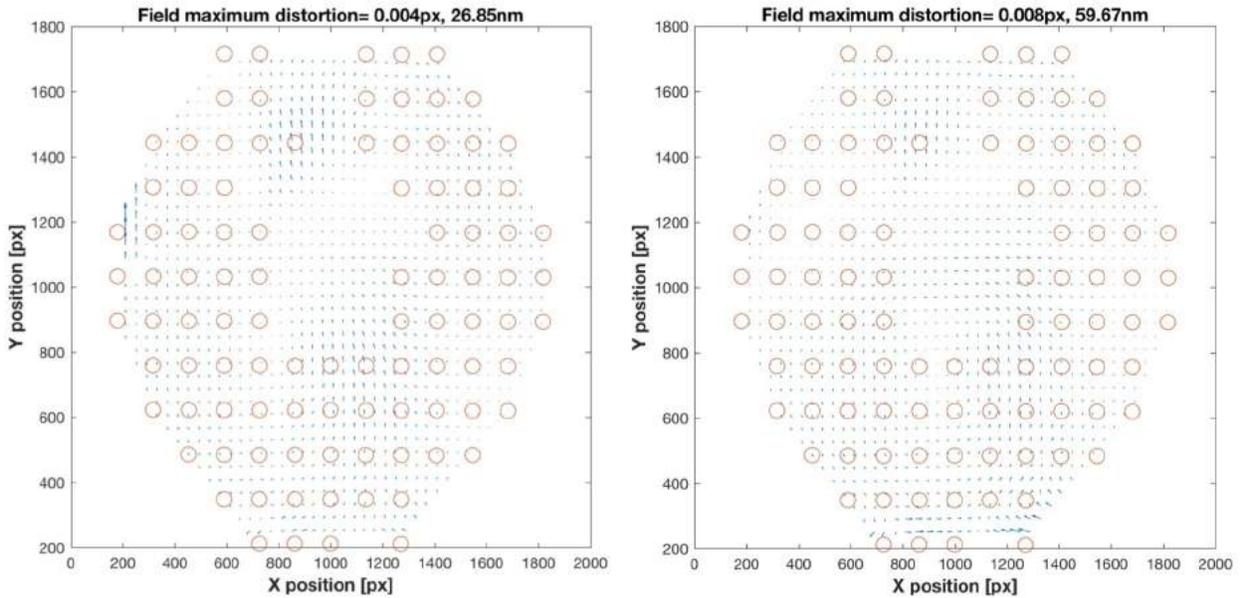

Figure 5.7: Sample of two different high-order distortion maps showing the correction applied to the stars in the field. The number on the top of each plots show the magnitude of the largest distortion vector in each image. Note that the distortion values are in the order of 5 milli pixel.

### 5.3 Direct imaging results

The goal of milestone #2 is to prove that high-contrast imaging techniques, including coronagraphs and wavefront control, remain operational and unaffected by the addition of a DP upstream of the coronagraph. Also, successful completion of Milestone #2, allows testing a full end-to-end system demonstration. The system layout and the coronagraph described in section 3.3 were used to perform the following tests.

Monochromatic laser light at 655nm was fed through the same optical path utilized by the astrometry instrument shown in the optical layout of Fig 3.1. The pick-off mirror extracts the high-contrast imaging FoV and feeds the coronagraph. For each run we used the following procedure:

- Align the PSF using the pre-PIAA tip/tilt mirror to remove any residual off-axis terms
- Calibrate plate scale and the PSF center position sending sine waves on the DM
- Insert the FPO rejecting star light to the PIAA LOWFS
- Image the rejected star light onto the LOWFS camera, then turn on the PID control loop
- Increase laser power to create visible speckle
- Execute speckle nulling to create the DZ

Typical residual rejected PSF jitter for the X and Y-axis was on the order of $1.4 \times 10^{-2} \lambda/D$ and $6.0 \times 10^{-3} \lambda/D$ RMS after drift control. The control loop was running at 1Hz to control drift, but not to suppress vibrations. The bench temperature was stabilized to +/-20mK during the run. A typical run would require a few hundred iterations over 2 days starting from a flat DM (due to using speckle nulling and not having a good model for EFC).

The contrast floor was improving every run as we learned more about the system and small improvements were performed after analyzing the data. For data set 1, we achieved a raw contrast floor of $4.78 \times 10^{-7}$ as shown in Fig. 5.8 left. During this run we discovered that vibrations and acoustic effects caused the by the Air Conditioning (AC) on/off cycle, which happens every 15 minutes approximately, slightly modify the speckle pattern. This prevented the speckle-nulling algorithm from improving the contrast further.

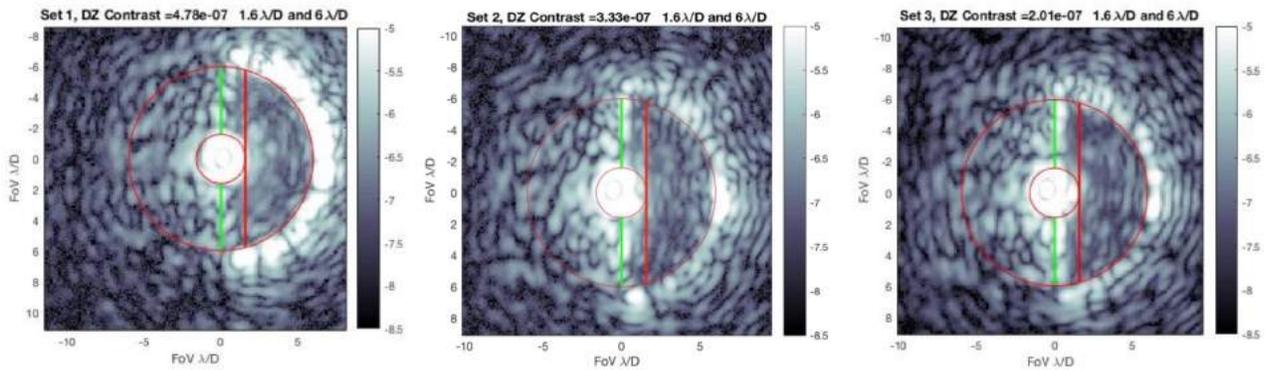

Figure 5.8: Images of the DZ achieved for data set 1, 2, and 3. All of them meet milestone #2.

Before starting data set 2 we modified the algorithm such that it will pause corrections when the LOWFS RMS increases beyond a certain threshold indicative of the AC starting. This approach allowed us to reach an average contrast of $3.33 \times 10^{-7}$ as shown in Fig 5.8 center, and reduce the variance as demonstrated by the blue line of Fig 5.9. Finally, for data set 3 we turned off the AC and hot-biased the optical bench to 24°C in order to have enough thermal control authority given the higher ambient temperature. The result was even better -- we were able to achieve $2.01 \times 10^{-7}$ from 1.6 to $6.0\lambda/D$, as shown in Fig 5.8 right, which is more than a factor of 2 deeper than the milestone contrast requirement of $5.0 \times 10^{-7}$.

The outer working angle is limited by the size of the C-shape of the occulter but is not a fundamental problem of the WFC algorithm or contamination caused by the DP. The example shown in Fig. 5.8 (right) is the best contrast in set 3, with 2.01x10$^{-7}$ contrast from 1.6 to 6.0λ/D. At this contrast level, there is no indication of impact caused by the DP. Higher contrasts are difficult to obtain due to effects from the PIAA lenses, mainly ghost reflections between lenses' flat surfaces. For each run, after the contrast floor was reached, we plotted the contrast of 25 consecutive frames to visualize contrast stability. The result is shown in Fig. 5.9. The instability of set 1 shown in black is the result of the AC cycle. For set 2, shown in blue, the control loop was stopped every time the AC started, and for set 3, shown in red, the AC was continuously off.

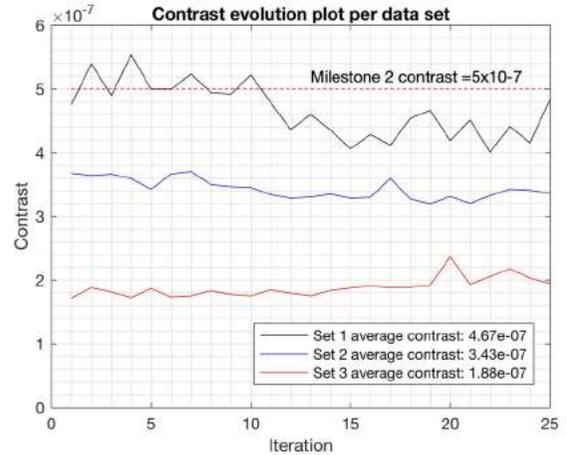

Figure 5.9: Contrast stability for each data set

To further explore the possibility of IWA contamination, we performed a basic post processing of the images. After the best contrast was achieved, the amplitude of the speckle nulling correction was dialed down to almost zero to maintain the configuration's stability and the first 25 images were averaged to create a reference frame. This frame was subtracted to the best contrast image in the set resulting in a DZ with an average contrast of 2.72x10$^{-8}$ (Fig 5.10) in which we could verify there is no DP impact down to a deeper contrast that in previous work [Bendek el al. 2013b]. Note that this simple post processing approach allowed us to gain a factor of about 10. This type of post-processing would reveal a planet if it was on different locations in the 25-image sequence, as would be the case, for example, in ADI (angular difference imaging),

Table 6.2: Milestone #2 run results, raw and post process contrast

| Milestone #2 | Working angle | | Raw contrast | Post process contrast |
|---|---|---|---|---|
| | IWA | OWA | | |
| Run 1 | | | 4.78x10$^{-7}$ | 2.81x10$^{-8}$ |
| Run 2 | 1.6 | 6.0 | 3.33x10$^{-7}$ | 9.11x10$^{-9}$ |
| Run 3 | | | 2.01x10$^{-7}$ | 4.79x10$^{-9}$ |

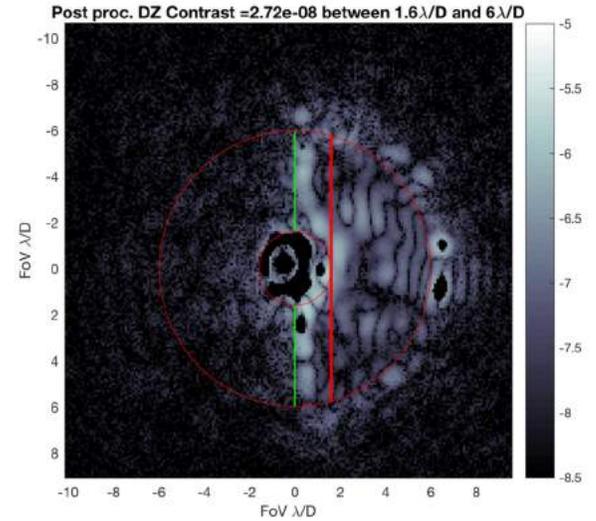

Figure 5.10: High contrast imaging example after performing average subtraction.

## 5.4 Coronagraph working angle calibration

We calibrated the coronagraph IWA with respect to the source located *before* the telescopes to avoid confusions caused the by magnification change induced by PIAA. We used two independent calibrations and also a PIAA fitting model verification.

*Approach 1: Calibrate source motion with respect to pre PIAA tip tilt*

This approach allows calibrating the working angle without any ambiguity associated the optics and magnifications that occur between PIAA and the sources. The procedure is as follows:

- Apply a 1 λ/D (8.44") tilt on the pre PIAA beam, where D is the telescope off-axis aperture. Since we used the pre PIAA fold mirror to modify the angle of the beam, only half of the angle (4.22") is applied to the Thorlabs Piezo mount KC1-PZ where the mirror is mounted. We calibrated the mount tilt rate to be 0.20"/V, so we applied 21.1V to obtain 1 λ/D beam tilt.

- Translate the light source position until the tilt applied before PIAA is fully compensated according to a LOWFS centroid measurement (which was empirically determined to be 58.4um +/- 10%. This enabled us to translate the fiber source by known amounts of off-axis displacements in units of l/D.

- Translate the fiber by known amounts of l/D and measure the energy of the resulting PSF on the science camera. The IWA is defined as the displacement at which the off-axis PSF energy is 50% of its large-angle asymptotic value. This was measured at 90.3$\mu$m fiber displacement, as a result the IWA = 1.55 λ/D on sky (see Fig. 5.11 top row)

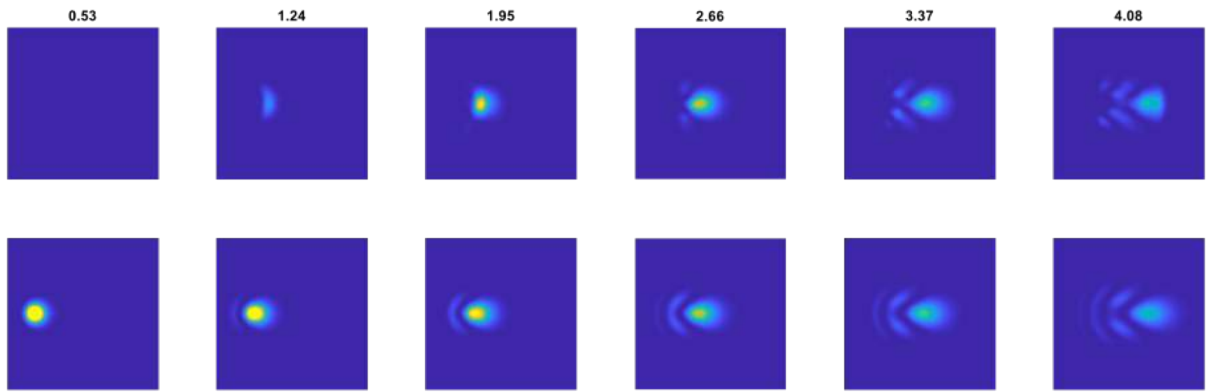

Figure 5.11: Sequence of PSF images as function of off-axis angle in terms of λ/D. The top row shows real data and the bottom row shows a simulation for the system. The real data is cropped on the left and the right by the C-Shape mask. This sequence was also used to determine the off-axis angle to obtain 50% of the PSF flux, thus defining the IWA.

**Approach 2: Calculation of pre PIAA beam tilt based on front end raytracing**

- We used the ZEMAX optics model to relate the light source motion in $\mu$m to λ/D angle before PIAA.
- According to our system model, 52.6$\mu$m is equivalent to λ/D. However the error is larger than in method 1 because of limited knowledge of beam expander configuration, which was re-configured in-situ, therefore no precise model is available.
- Using the 50% throughput definition of IWA, the result is IWA = 90.3/52.6=1.71 λ/D on sky.

**Sanity check: Off-axis PIAA model PSF fit test**
- Displace the source by incremental amounts, and take images the PSF for different locations.
- Convert linear translations to λ/D using method 1.
- Compute theoretical PSFs and fit theoretical PSFs to images.

- The final fit result consistent with the results computed using approach 1 and 2. Plots of the PSF fits are shown in Figure 5.12. The images show that an IWA=1.6 is more consistent with this sanity check than, say, IWA=1.28.

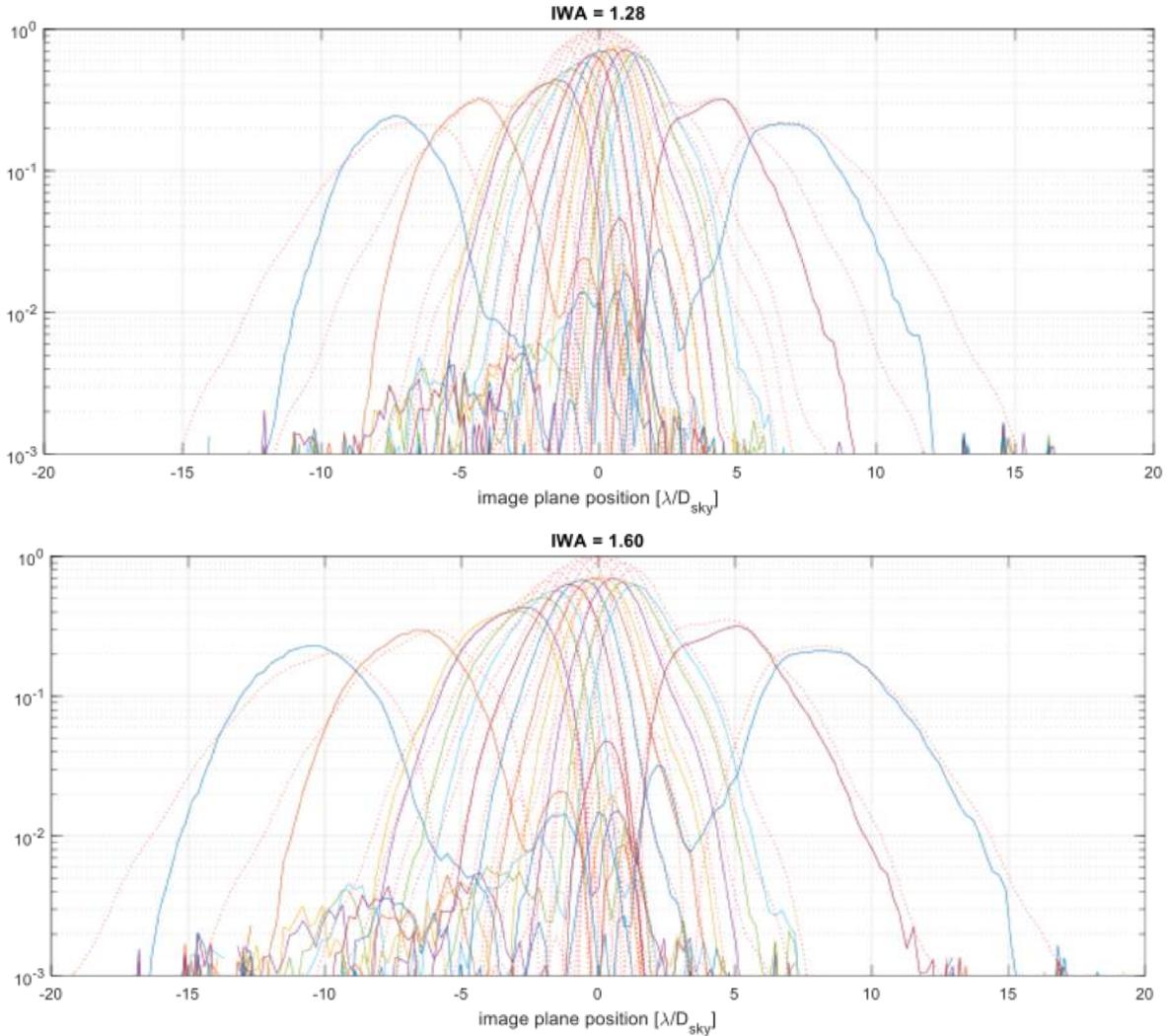

Figure 5.12: Comparison of off-axis PSF for a 1.28 λ/D IWA coronagraph (top) and for 1.6 λ/D (bottom). Thick lines are measurements and thin lines are models. Aberrations caused upstream of PIAA has deformed slightly the PSF resulting on a loose fit. However, the 1.6 λ/D fit is the closest one to the model.

**Conclusion:**

The values of 1.55 λ/D and 1.71 λ/D obtained using approach 1 and 2 have an average of 1.63λ/D. Both samples are within 5% of the average, and the sanity check PSF fit is consistent with a system with the 1.60λ/D IWA value. Therefore, we conclude that the IWA for the system using the definition of 50% of the PSF throughput is 1.63 λ/D. Since the milestones have been defined only with one decimal point we consider reasonable to round the calibrated IWA to 1.6 λ/D.

### 5.5 Coronagraph contrast calibration

To calibrate the contrast of the dark zone we recorded exposure times and laser power for every exposure. We calibrated the stability of the laser power when it was operated at constant power and temperature and we discovered there could be power fluctuations up to 15% without any setting change. We decided to calibrate the contrast always using two speckles on the same image to provide resilience to laser brightness oscillations. In addition an approximation of the contrast was obtained in real time from the intensity in the LOWFS camera.

The contrast calibration procedure relies on measuring the relative brightness between different speckles in the same image. The dynamic range is increased every time a dimmer image is taken, hence relating the brightness of a measured speckle with a new one that was saturated on the previous image, similar to how High Dynamic Range (HDR) imaging works. By repeating this process multiple times, it is possible to obtain the dark zone contrast with respect to the host star PSF.

Let's define a typical dark zone residual speckle brightness as $s_1$, and $s_2$ as the brightest non-saturated speckle that can be measured in the same image, then we can define the speckle brightness ratio in the same image as:

$$R_1 = s_1/s_2$$

For the next image, we reduce the light source intensity or exposure time in order to measure the brightness of brighter speckles, but always maintaining enough flux to be able to measure, with SNR 5 or more, the brightest speckle of the previous image. So, for any given image we compute the brightness ratio as:

$$R_i = s_i/s_{i+1}$$

We repeat this procedure until the PSF core is not saturated and we are able to compute the speckle to PSF brightness ratio in the same fashion:

$$R_{sp\_psf} = s_i/s_{psf}$$

Finally, the dark zone speckle contrast can be written as:

$$C_{speckle} = R_i \cdots R_i \cdots R_{sp\_psf}$$

Then we can express the average contrast of the dark zone as:

$$C_{DZ} = \frac{I_{DZ}}{C_{speckle}}$$

Where $I_{DZ}$ is the average brightness of the Dark Zone pixels and $C_{DZ}$ is the Dark Zone contrast relevant for the milestone. To increase the technique robustness we measured several pairs of speckles per image to avoid localized noise affecting the measurements.

After the approximate contrast indicated by the LOWFS camera was better than *5x10$^{-7}$* and it was stable for several iterations, we stopped the wavefront control and measured the flux of several bright speckles. Then, we reduced the power until the bright speckles were dim. On every iteration, we used at least a bright and dim speckle to be able to connect the contrast ratio from image to image without relying on the laser power information.

# 6 Milestones verification

## 6.1 Milestone 1

**Milestone #1 definition:**
**Broadband medium fidelity imaging astrometry demonstration**

*Demonstrate $2.4 \times 10^{-4}$ $\lambda/D$ astrometric accuracy per axis performing a null result test. The laboratory work will be carried out in broadband spectrum covering wavelengths from 450 to 650nm using an aperture pupil (D) equal or larger than 16mm.*

The "angular separations" are defined in terms of the source wavelength $\lambda$, and the diameter $D$ of the aperture on the DP, which is the pupil-defining element of the imaging astrometry camera. For this milestone, a DP simulates the telescope primary and pupil and it was illuminated by an array of broadband point sources forming f/25 to f/50 beams when they reach the pupil.

### 6.1.1 The astrometry milestone success criteria

The success criterion is met when the standard deviation of a set of N (where N>10) single observations is < $2.4 \times 10^{-4}$ $\lambda/D$. Rationale: This accuracy corresponds to a medium to high-fidelity demonstration of the technique, meeting one of the criteria to reach TRL-4.

### 6.1.2 Milestone #1 success criteria verification

The success criteria was satisfied and exceeded by ample margin for the 3 different data runs. Each run was 48hrs long, with epochs separated every 4 hours completing 12 epochs per data set. Table 6.1 summarizes the results.

Table 6.1: Milestone #1 verification matrix

| Milestone #1 | Null Test RMS [$\lambda/D$] | | Milestone Goal [$\lambda/D$] | Times better than milestone | Milestone #1 met? |
|---|---|---|---|---|---|
| | X-axis | Y-axis | | | |
| Run 1 | $8.93 \times 10^{-5}$ | $5.45 \times 10^{-5}$ | $2.4 \times 10^{-4}$ | 3.3 | Yes |
| Run 2 | $9.62 \times 10^{-5}$ | $3.02 \times 10^{-5}$ | | 3.8 | Yes |
| Run 3 | $4.07 \times 10^{-5}$ | $3.43 \times 10^{-5}$ | | 6.4 | Yes |

## 6.2 Milestone 2

**Milestone #2 definition:**
**Broadband medium fidelity simultaneous imaging astrometry and high-contrast imaging**

*Demonstration of milestone #1, and performing high-contrast imaging achieving $5 \times 10^{-7}$ raw contrast between 1.6 and $6\lambda/D$ by a single instrument, which shares the optical path, from the source to the coronagraphic and astrometry FoV separation. The ability of achieving $5 \times 10^{-7}$ raw contrast will be considered as proof of no contamination of the IWA.*

### 6.2.1 Success criteria

The following are the required elements of the milestone demonstration. Each element includes a brief rationale.

### 6.2.2 Direct imaging milestone success criteria

A mean contrast metric of $5 \times 10^{-7}$ or smaller shall be achieved in a 1.6 to 6 $\lambda/D$ dark zone. *Rationale: This provides evidence that medium to high fidelity demonstration of direct imaging is* *compatible with a DP telescope and the astrometry measurements. Also contrast of $5x10^{-7}$ is in the regime where useful science can be achieved.*

### 6.2.3 Milestone #2 Success criteria verification

For each wavefront control run, we started from the flat deformable mirror voltages in order to satisfy the milestone #2 condition that states:

*"The wavefront control system software reset between data sets ensures that the three data sets can be considered as independent and do not represent an unusually good configuration that cannot be reproduced. For each demonstration the DM will begin from a "flat" setting."*

The contrast requirement specified by the success criteria was satisfied with margin for the three runs.

Table 6.2: Milestone #2 verification matrix

| Milestone #2 | Working angle | | Raw contrast | Milestone #2 requirement | Milestone #2 met? |
|---|---|---|---|---|---|
| | IWA | OWA | | | |
| Run 1 | | | $4.78x10^{-7}$ | | Yes |
| Run 2 | 1.6 | 6.0 | $3.33x10^{-7}$ | $5.0x10^{-7}$ | Yes |
| Run 3 | | | $2.01x10^{-7}$ | | Yes |

# 7   Conclusions

The scientific importance of measuring planet masses is increasing as the exoplanet community focuses on exoplanet characterization in order to answer NASA's science plan and 30-year road map question of whether potentially habitable exoplanets exist and whether they can harbor life. Masses are particularly important in the study of earth-like planets to distinguish them from small Neptunes or water worlds and to predict atmospheric retention and improve transmission spectroscopy models.

The work carried out during this technology development grant has allowed us to demonstrate and advance the DP technology that can be used to calibrate optical system distortions, and hence, improve the imaging stellar astrometry accuracy to sub-uas levels. Such accuracy could enable measuring masses of earth-like planets around nearby stars. In addition, we have shown that the DP technology is technically fully compatible, and scientifically synergistic with a coronagraph.

We have met milestone #1 with a comfortable margin. The average accuracy obtained during all the runs is $5.75 \times 10^{-5}$ L/D, which is 4 times more precise than the milestone requirement, and equivalent to $2.5\mu as$ on 2.4m telescope, or $1.5\mu as$ for a 4m telescope working in visible band. These results show the potential of this technique to enable future exoplanet mission to detect and measure masses of earth-like planets around nearby stars. This will bring a significant benefit to the astronomy community.

We met milestone #2 and demonstrated that it is possible feed a coronagraph with a telescope equipped with a DP and achieve high-contrast imaging. We did three different high-contrast imaging runs and met milestone #2 of $5 \times 10^{-7}$ raw contrast for all of them. On average, we obtained $3.33 \times 10^{-7}$ raw contrast considering all data sets together. This result is 35% better than the milestone #2 requirement. We validated the stability of the high-contrast region by averaging frames and subtracting the average from single frames, which resulted in contrast improvement of almost one order of magnitude, reaching $2.72 \times 10^{-8}$ contrast.

*Lessons Learned*

We learned many aspects of the experiment that we would like to improve and share with the community that would like to embark on similar investigations. We found that the residual star jitter noise after applying high-order distortion correction is mostly caused by lack of ***detector characterization and calibration***. Pixel response and intra-pixel Quantum Efficiency variations bias the star position when the PSF is moved a fraction of a pixel, which is the case during these experiments. Performing detector calibration would allow us to mitigate this effect.

We realized that ***clean room environmental variables*** are critical and often are poorly characterized. We placed the astrometry bench at the ACE laboratory, which is a class 100,000 clean room with thermal control stable down to 1°F. We designed a thermal enclosure to filter those temperature variations and stabilize at hot-biased equilibrium temperature. However, soon we discovered that the AC fan was transmitting vibrations through the walls to the bench, and also that there were acoustic perturbations caused by the fan and the ventilation duct blower. For milestone #2 this issue became our main limiting factor to achieve the contrast required. We solved the problem by connecting the clean room with the one next door and stopping the AC for the ACE lab. This enabled us to improve the contrast by a factor of 10.

Another important aspect for proper operation of any wavefront control algorithm and to calibrate contrast as well is the ***light source power stability***. Closed loop power control based on current is very sensitive to temperature and it could mislead the contrast calibration. An absolute and independent laser power measurement using the light leak from a mirror seems the best option to solve this problem.

*Future Work*

An approach that would allow us to improve the accuracy without calibrating a detector is to roll the telescope with its bore sight aligned with the host star. The resulting image would have the spikes unchanged. However, the stars will describe arcs centered on the target star, spreading the light of those point sources across many pixels and therefore rapidly reduce the impact of detector miscalibration and intra-pixel sensitivity variations. Since rotating the telescopes on the bench is cumbersome, we mounted the light source on a rotary stage, which is equivalent to performing a telescope roll. The hardware is installed and we hope to continue the experiment if there is interest in the community.

Further optimization of the algorithm is being considered. This includes masking the stars/spikes to calculate their distortion estimates separately, creating an adaptively-sized interpolation kernel, and time-varying kernel size that can change while interpolating.

On the coronagraph side, we will use the optical bench to continue research and algorithm development of Multi-Star Wavefront Control (Thomas et al. 2015, Sirbu et al. 2017), where we dig a dark zone around two nearby light sources, simulating binary stars on the sky. This technique requires a novel wavefront control algorithm that uses different zones of the DM to null speckles of the two stars. Currently, we are installing a fiber bundle with fibers separated by every 200μm allowing us to simulate two or more stars.

Finally, we would like to invite the community to use our facility and/or send experiments of interest in the context of astrometry and direct imaging. For example it would be fairly easy to replace the PIAA coronagraph for a Vortex or a Vector APP.

# 9 List of Acronyms

| | |
|---|---|
| AC | Air Conditioning |
| ACE | Ames Coronagraph Experiment |
| AD | Astrometry Demonstration |
| ARC | Ames Research Center |
| BMC | Boston Micromachines Corporation |
| DM | Deformable Mirror |
| DP | Diffractive Pupil |
| DZ | Dark Zone |
| EFC | Electric Field Conjugation |
| FPO | Focal plane occulter |
| FFOV | Full Field of View |
| FOV | Field Of View |
| FWHM | Full Width Half Maximum |
| HFOV | Half Field Of View |
| IWA | Inner Working Angle |
| LOWFS | Low Order Wavefront Sensor |
| NWNH | New Worlds New Horizons in Astronomy and Astrophysics |
| OD | Optical Density |
| PIAA | Phase Induced Amplitude Apodization |
| PID | Proportional Integral Derivative |
| PSF | Point Spread Function |
| PV | Peak to Valley |
| RMS | Root Mean Square |
| RON | Read Out Noise |
| SN | Speckle Nulling |
| SNR | Signal to Noise Ratio |
| TDEM | Technology Development for Exoplanet Missions |
| WFC | Wavefront Control |

# 10 Appendix

External reviewer comments

**Context**

Dr. Mike Shao served as an external reviewer assisting the TAC to evaluate the TDEM work. The TAC considered the questions and answers valuable; therefore, we decided to capture it on this Appendix.

**Mike Shao comments to TDEM final report** – *Answers in Italic and highlighted in yellow*

Questions and comments on Bendek TDEM final report

**Dr. Shao:**
"Overall comments. I agree with the PI that astrometry is a very important part of exoplanet search and characterization. Also the diffraction pupil is an innovative way to address one of two (and only two) major systematic error sources. Photons from the target and reference stars enter the telescope, and are detected by the detector. Both the telescope and detector have imperfections and must be calibrated.

This demonstration is a first major step and the following questions regarding scaling address how this would scale to a mission. But it is appropriate in this 1st step to demonstrate the technology under laboratory conditions and at a later date scale future experiments to a mission.

1) The scaling questions are:

The FOV of the experiment is 19 arcmin, but in a mission the 20mm pupil will be replace by a 1m telescope or 4m telescope or 12m telescope. What is its FOV?"

*PI Answer:*
*If we assume a constant number of stars of magnitude X in the FoV, then the astrometric accuracy will scale as L/D for the telescope size. Regarding the FoV, the laboratory performance demonstration will not change as long as the number of background stars that are 10 magnitudes dimmer is constant. For the laboratory experiment we are using about 120 stars, which on average are 10 mag dimmer than the host star and their PV brightness is contained within 1 mag range.*

*Whether this configuration is realistic or not will depend of how bright is the target star. For example: If our target is a magnitude 3 star, we will get 136 mv=13 stars (source* http://spacemath.gsfc.nasa.gov/stars*). As we go fainter this situation becomes more favorable. If our target star is mv=5, then we will find 716 m=15 stars in the same FoV. Therefore, it is possible to reduce the FFoV to less than half (FFoV =16.2 arcmin) and obtain the same performance assuming that the telescope is large enough that there is no photon noise penalty.*

*\*NOTE: Our experiment has a FFoV of 37.2 arcmin and the background stars are on average 10 magnitudes dimmer.*
**Dr. Shao**

"This is tied to question 2, which is, in that FOV, what reference stars does one expect to see?"

*PI Answer:*
*For target mv=3, => 136 mv=13 stars*
*For target mv=5, => 716 mv=15 stars*

**Dr. Shao**
"A related question is in the experiment the target star is 8 mag brighter than the reference stars, is this realistic given the FOV of a mission? (Fig 3.3 says 8 mag but the text immediately below says 1.4e4 brighter ~9 mag. At this stage we're not worried about 1mag, but rather if there may be issues later on when implementing this on a mission. But even if it's there is a disconnect when tracking to a mission, the technology development is an important first step. A future experiment can address the traceability."

*PI Answer:*
*Thanks for the comment. I calibrated the central to background brightness star ratio. The result is 10.5 magnitude difference. For purpose of the text I will round at 10 magnitude difference.*

**Dr. Shao**
"The large mag diff between target and ref stars is necessary because the diffracted dots/streaks are quite faint. If the ref stars are very bright they will saturate before the diff dots/spikes get above the read noise.

2) A second mission related question is:

The DP is a way to monitor changes in field distortion while taking science data. When used with a normal telescope, like HST, it would seem that one can expect field distortion to change at the uas level on a time scale of minutes to hours. But a coronagraphic telescope needs much higher stability.

Calibrating distortion using stars is the more traditional approach, but a dense field of stars needed to calibrate to the uas level in general is not available for most explanet targets. And then one has to assume the telescope is stable as it move from the calibrator field to the target field. Missions like WFIRST have done detailed thermal modeling, would field distortion calibration be possible using only reference stars? But like other mission scaling questions this is not a primary focus of this TDEM, which aims to develop/test the technology."

*PI Answer:*
*I think that field distortion calibration using reference stars is not reliable enough to ensure astrometry performance and science return. Finding a star cluster for which we know the stars positions to an accuracy of 1uas and it is very close to our scientific target is very unlikely. In addition, the telescope will deform as the observation takes place. The DP offers continuous calibration on the target that calibration stars cannot offer.*

**Dr. Shao**
"3) Astrometry experiment.

TDEM milestone #1. Null test. Clearly meeting this null test is a necessary condition towards demonstrating the ability to make uas measurements or 1e-4 l/D astrometry. But this null experiment by itself doesn't allow the demonstration that the DP concept or the data analysis actually works.

In an astrometric measurement of stars, through a telescope, the telescope changes slightly during a measurement (lasting a hour or more) and also changes over a time scale of months/years as the same target is observed dozens of times to measure the reflex motion of the star.

Even if the telescope were perfectly stable differential stellar aberration will change the angle between the target star and reference star as the Spacecraft goes around the Sun. This was a very large effect for Kepler, which had a very large FOV, but for even a 1/2 deg FOV (0.01 rad) this effect is 200 mas (multiple pixels and multiple lambda/D). Of course the purpose of DP is to measure changes in the telescope's distortion.

In the experiment, the tip/tilt was controlled to keep the star field static to ~10% of a pixel. For a 4m telescope, (and 6 pix/(lam/D)) this is a motion of lam/(D*60) ~ 0.4 milliarcsec. Diff stellar aberration can move stars by ~200mas.

The point is that let's say the DP doesn't work or the data analysis is incorrect by a factor of 2. In a static experiment, nothing is changing and the factor of 2 error is never exposed. Now nothing is perfectly static, but the 1/10 pixel motion in the experiment is a few 100 times smaller than the motion due to differential stellar aberration, for a perfectly stable telescope with perfect pointing."

*PI Answer:*
*Thank you for pointing out the possibility of multiplicative errors. We calibrated and demonstrated resilience of the algorithm to multiplicative errors on section 4.2 and 4.3. We do acknowledge that those tests can only validate the algorithm and not the resilience of the DP physics itself to multiplicative errors, we don't foresee any reason why the DP physics can be compromised.*

*Our experiment has been executed in a similar fashion that a space mission would operate. We agree with the reviewer that errors on the Kepler mission are larger that stability that we provide on our laboratory. However, as the reviewer mentions in question 2, the requirements for future exoplanet missions will be much tighter than Kepler making our laboratory environment reasonable.*

*For example, during a data set, it was possible to turn off the tip/tilt Piezo controllers moving away the star from the FoV between epochs measurements. As soon as the voltage was restored on the controller, the star will get back with in the FoV, the PID control loop was started bringing the star within 10% of a pixel of the position it was before, and in the correct position for the next epoch measurement.*

*The differential stellar aberration is of course very important but beyond the scope of what we considered in this work. It would be [I think] indistinguishable from internal telescope or instrument aberrations, but it is known a priori to many significant figures and can be subtracted out. This*

*should mitigate this effect, but it is possible that the residual after subtraction may be a limit of DP astrometry for planets close to 1-year period and its harmonics.*

**Dr. Shao**
"4) Data reduction algorithms

The data analysis approach presented is highly nonstandard, in that it does not follow the approach used by astronomers doing astrometry on the ground or with for example HST.

The approach in data analysis looks at the gradient of the image. This approach only works for small displacements, motions small compared to lambda/D. The pixel scale is 6 pixels per lambda/D and the test motion was 1/2 pixel so the test situation satisfies the condition of "small" motion. It measures "relative" distortion (the change in distortion between two epochs where the motions are small), rather than absolute distortion.

But the general astronomical situation has much larger motions. Differential stellar aberration can produce motions of 0.2 arcsec over a 0.5deg FOV. Many target stars for Exo-Earth detection are nearby, (< 10pc) and would have proper motions of 0.5 to even 1 arcsec over 5 years. Expected motions of stars are much larger than 0.025 arcsec lambda/D for a 4m Habex telescope.

Eventually a modified algorithm is needed that can measure the macroscopic motions of the target star, 100's of lambda/D."

*PI Answer:*
*We agree with the reviewer that the algorithm is optimized for small relative motions and relative distortions, because sensitivity to very small distortions and astrometric signals was what we perceived that to be the most challenging aspect of the technology. However, we don't expect larger deformations of the spikes on a stable space telescope. We are confident that it is possible to position the host star back within half a pixel accuracy using a high accuracy pointing control loop such as the WFIRST LOWFS control. In that case, the spikes will appear with the 0.5px threshold over the whole image unless large optical distortions are occurring, which we don't expect on a stable space telescope.*

*We do agree that background stars will have much larger motions than 0.5px, however the position of background stars is measured with traditional centroiding and therefore is resilient to large motions. We obtain the distortion map from the spikes and then we can apply that map to any location where the stars are found in the next epoch enabling measuring large background stars motion.*

**Dr. Shao**
"5) General comments on null tests

There are several different types of null tests. Totally null test. This where nothing moves. The problem with this test is that if the DP approach or the data analysis algorithm doesn't work it will still pass the totally null test.

Stationary target test. In this test the relative positions of the target star and ref stars do not move, but the whole field is move across the detector/image plane by a few 100 lambda/D, typical of the motions one would expect real stars to move. It would actually be useful to do this test without a coronagraph present. One can then do "normal" astrometry (fitting PSF's to the images", as well as DP (diff pupil) astrometry making use of the spikes. One can then see the field distortion error in "normal" astrometry and a much-reduced amount in DP astrometry.

Fully articulated test. In this test the star field is moved over the focal plane by many 100's of lambda/D, but in addition, the target star is moved relative to the reference stars. The motion of the target star should be consistent with both differential stellar aberration and proper motion of nearby exoplanet target stars."

*PI Answer:*
*It would be an excellent next step to implement a fully articulated test, however it is outside the scope and cost of this TDEM effort. I would like to explore with the reviewers and the community how to implement this test in the future.*

**Dr. Shao**
"6) Diff Pupil and photon noise and other uses

The baseline DP blocks 1% of the light from a star and diffracts that light into 1000's of spikes. The result is that the spike even from a 5 magnitudes target are very dim and the photon noise from the spikes can be a serious noise source. In more traditional astrometry, photon noise of the reference stars is the limiting "random" noise source. But while this may or may not be a serious issue for in flight use, the diffractive pupil can be an extremely useful tool for testing uas distortion of telescope while they are in the lab prior to launch.

In traditional astrometry using reference stars that solve for polynomial models of the distortion, it's difficult to find fields of view with 1000 or perhaps even 10,000 moderately bright stars. The DP supplies these in the lab with well-separated fake stars. There are dense starfields, but often they are so dense that stars will merge. While one would expect the field distortion of a telescope in the lab to change significantly after it's in space, we expect most of the change to be low spatial freq. If this is true Diff Pupil calibration on the ground would simplify distortion calibration in space even if we don't use the DP in space."

*PI Answer:*
*We appreciate those ideas and comments.*

**Dr. Shao**

"7) Last comments

Given this is the final report, there may not be enough time or $ resources to do test #2 the stationary target test. But if there is a follow on TDEM or some additional resources can be found, I this this would be a very useful addition to that large body of work already done."

*PI Answer:*

*Completely agree. I am happy to explore options to continue the experiment. Maybe we can apply to just build the ultra precise fully articulated astrometry source as an add-on to the current astrometry bench.*